\documentclass[5p]{elsarticle}

\makeatletter
\def\ps@pprintTitle{%
  \let\@oddhead\@empty
  \let\@evenhead\@empty
  \let\@oddfoot\@empty
  \let\@evenfoot\@empty
}
\makeatother

\usepackage[title]{appendix}

\biboptions{numbers,sort&compress}

\usepackage{libertine}
\usepackage{libertinust1math}
\usepackage{geometry}
\usepackage{amsmath}
\usepackage{bbold}
\usepackage{graphicx}
\usepackage{eurosym}
\usepackage{mathtools}
\usepackage{url}
\usepackage{booktabs}
\usepackage{epstopdf}
\usepackage{xfrac}
\usepackage{tabularx}
\usepackage{bm}
\usepackage{subcaption}
\usepackage{longtable}
\usepackage{multirow}
\usepackage{threeparttable}
\usepackage[export]{adjustbox}
\usepackage[version=4]{mhchem}
\usepackage[colorlinks]{hyperref}
\usepackage[nameinlink,sort&compress,capitalise]{cleveref}
\usepackage[leftcaption,raggedright]{sidecap}
\usepackage[prependcaption,textsize=footnotesize]{todonotes}
\usepackage{blindtext}
\usepackage{url}

\usepackage{xr}
\usepackage{lipsum}

\graphicspath{
    {../workflow/notebooks/},
    {../workflow/pypsa-eur/resources/},
    {../workflow/pypsa-eur/results/},
}

\usepackage{siunitx}
\DeclareSIUnit\year{a}
\DeclareSIUnit{\tco}{t_{\ce{CO2}}}
\DeclareSIUnit{\sieuro}{\mbox{\euro}}
\DeclareSIUnit{\twh}{{\tera\watt\hour}}
\DeclareSIUnit{\mwh}{{\mega\watt\hour}}
\DeclareSIUnit{\kwh}{{\kilo\watt\hour}}

\renewcommand{\ttdefault}{\sfdefault}

\newcommand{\ubar}[1]{\text{\b{$#1$}}}

\begin{document}

\begin{frontmatter}

	\title{Price Formation in a Highly-Renewable, Sector-Coupled Energy System}

	\author[tub]{Julian Geis\corref{correspondingauthor}}
	\ead{j.geis@tu-berlin.de}
    \author[tub]{Fabian Neumann}
    \author[tub]{Michael Lindner}
    \author[iee]{Philipp Härtel}
	\author[tub]{Tom Brown}

	\address[tub]{Department of Digital Transformation in Energy Systems, Institute of Energy Technology,\\Technische Universität Berlin, Fakultät III, Einsteinufer 25 (TA 8), 10587 Berlin, Germany}
	\address[iee]{Fraunhofer Institute for Energy Economics and Energy System Technology, Fraunhofer IEE, Kassel, Germany}
	\begin{abstract}
As variable renewable energy increases and more demand is electrified, we expect price formation in wholesale electricity markets to transition from being dominated by fossil fuel generators to being dominated by the opportunity costs of storage and demand management. 
In order to analyse this transition, we introduce a new method to investigate price formation based on a mapping from the dual variables of the energy system optimisation problem to the bids and asks of electricity suppliers and consumers. This allows us to build the full supply and demand curves in each hour. We use this method to analyse price formation in a sector-coupled, climate-neutral energy system model for Germany, PyPSA-DE, with high temporal resolution and myopic foresight in 5-year steps from 2020 until full decarbonisation in 2045. 
We find a clear transition from distinct price levels, corresponding to fossil fuels, to a smoother price curve set by variable renewable energy sources, batteries and electrolysis. Despite higher price volatility, the fully decarbonised system clears with non-zero prices in 75\% of all hours.
Our results suggest that flexibility and cross-sectoral demand bidding play a vital role in stabilising electricity prices in a climate-neutral future. These findings are highly relevant for guiding investment decisions and informing policy, particularly in support of dynamic pricing, the expansion of energy storage across multiple timescales, and the coordinated development of renewable and flexibility technologies. 	\end{abstract}


\end{frontmatter}

\section{Introduction}
\label{sec:intro}

\subsection{Motivation}

Electricity systems are being transformed from several different directions at once: zero-marginal-cost variable renewable energy sources (VRES) are starting to dominate supply in some markets; flexible demand from electrified transport and heating is increasing; and rising storage capacity is narrowing price differences through arbitrage. The literature focused on the power sector has examined how rising VRES increases zero-price hours, thereby lowering average prices and market values and making it harder for generators to recover their costs \cite{sensfus_merit-order_2008,ketterer_impact_2014,eu_comission_2023}. At the same time, studies that included the electrification of demand and sector-coupling have observed a somewhat countervailing trend: demand in other sectors extends the demand curve, meaning that demand is more often price setting. Böttger and Härtel used a heuristic in a sector-coupled model to identify times when specific technologies were price setting, incorporating technologies one at a time \cite{bottger_wholesale_2021}. However, the field has been missing analytic tools that allow to reconstruct the full supply and demand curves from all the opportunity costs of storage, demand-side management and other energy converting technologies. In this paper we build on \cite{brown_price_2025}  to introduce a general method for deriving the full bidding curves from optimisation models, enabling us to analyse in detail what is price setting, and how this changes over time.

\subsection{Literature}
Recent literature reveals different challenges that a future decarbonised electricity market faces concerning efficient market clearing and price formation.\newline
The rapid expansion of VRES electricity generation in the German and the worldwide energy mix has significant effects on the spot prices. The integration of VRES into the wholesale market increases the supply of low marginal cost electricity, displacing higher-cost generators in the merit order and thereby reducing the market clearing price. This so-called merit order effect of solar and wind is well documented in the literature especially for the German market \cite{hildmann_empirical_2015, cludius_merit_2014, sensfus_merit-order_2008}. 
\newline 
Without any countervailing measures, the market revenues for VRES are cannibalised as their market values decline with higher shares of wind and solar PV \cite{das_learning_2020, pena_cannibalization_2022, reichenberg_revenue_2023}. In periods of high wind and solar penetration, this additional electricity supply reduces the prices that these technologies receive. In a study based on a European market model, this cannibalisation effect is esimated to decrease the wind market value from 110\% of the average electricity price to 50-80\% at the transition from near-zero penetration to penetration of 30\% in the German market \cite{hirth_market_2013}. These trends are also found in a simulation of California's day-ahead wholesale electricity market \cite{lopez_prol_cannibalization_2020}. Hildmann et al. \cite{hildmann_empirical_2015} confirm these effects empirically, but argue that the power market would remain functional with regulatory adjustments and the incorporation of all VRES cost components. They claim that the marginal costs of VRES are positive due to wear and tear costs, concession tax, land lease costs, and forecast errors. Brown et al. \cite{brown_decreasing_2021} argue that this decline in revenues is a result of policy choices and not an intrinsic property of VRES. It only occurs in systems where technologies are forced into the market by subsidies or quotas. \newline 
A concern emerging from the reduction in the market values and cannibalisation of own profits is that VRES are no longer able to cover their investment costs \cite{mays_financial_2023, egli_renewable_2020, mcelroy_missing_2018, hogan_2017}.
Additional concerns relate to an increased price volatility and the frequency of zero or negative prices \cite{han_extremal_2024}. Mallapragada et al. \cite{mallapragada_electricity_2023} confirm for different scenarios of capacity expansion modelling in the US that the future system has more hours with very low and relatively high prices. In terms of cost recovery, this means that most of the profit is made in just a few hours when prices are high. Even the inclusion of cheap, long-duration storage does not fully mitigate the increase in near-zero electricity prices in their study.\newline 
\\

All these challenges raise major concerns, that the energy-only market (EOM) may no longer be functioning \cite{thomasen_will_2022}. Some studies suggest those challenges need to be addressed by a change in the market model towards capacity mechanisms. Taylor et al. \cite{taylor_power_2016} express concern that the current nodal pricing framework, as it is employed in e.g. North America, is no longer suitable for VRES with low marginal and high fixed costs. Blazquez et al. \cite{blazquez_renewable_2018} argue that the EOM necessitates the coexistence of fossil fuel-based generation. According to their study, the price setting of dispatchable power plants is essential for VRES to generate a profit margin. Without fossil generators, they elaborate, the price formation would be dominated by VRES, leading to a situation where the price remains zero most of the time.\newline
\\

Other studies which incorporate sector coupling, trade or flexible demand find different price and market behaviour. The electrification of e.g. land transport and space heating as well as the interaction with hydrogen prices provide additional flexibility. This can reduce the effect of price volatility and zero prices \cite{neumann_potential_2023, helisto_impact_2023-1}. Böttger and Härtel \citep{bottger_wholesale_2021} investigate the market clearing effect of a stylised market with cross-sectoral bidding, storage, and sector coupling. VRE market values are stabilised by cross-sectoral trading, storage, power-to-gas (PtG) and power-to-heat (PtH) technologies. This leads to moderate price duration curves and fewer zero prices. In contrast to the more forward-looking effects of sector coupling, their analysis of the European electricity system shows the additional stabilising role of increasing cross-border electricity trade, which is already observable in today’s markets. Bernath et al. investigate how different flexibility options provided by efficient sector coupling impact the market values of VRE \cite{bernath_impact_2021}. They find that the market values of VRE are especially increased by the use of electricity in district heating. 
Brown et al. \cite{brown_price_2025} argue that the problem of price singularity is caused by modelling with perfectly inelastic demand. In their model with short-term elasticity and only including wind and solar generators as well as batteries and hydrogen-based storage, they see a reduced fraction of zero-price hours from 90\% to about 30\%. This is consistent with the flexible contribution of sector-coupling technologies providing elasticity to the electricity market.\\
\newline
Some studies go further and identify the specific technologies that set prices in each hour. A study on the price setting of the German market reveals that in 2020, the major price-setting technologies were gas (30\%), coal+lignite (31\%) and pumped hydro (21\%) \cite{blume-werry_eyes_2021}. The Joint Research Centre of the European Union  \cite{gasparella2023merit} investigated price-setting dynamics in European electricity markets. They find that gas sets 55\% of the prices in 2022 and is modelled to continue on this level until 2030 across Europe. The planned increase in interconnection capacity will reduce cross-border price differences and stabilise prices. Korpås et al. \cite{korpas_optimality_2020} model a stylised system with linear supply functions. They find that all technologies cover their costs in equilibrium, even with high VRES and curtailment. Härtel et al. \cite{hartel_demystifying_2021} analyse a European model with 87.5\% CO$_2$ reduction, showing that price-setting technologies include direct resistive and heat pump heating, with fewer than 15\% zero- or negative-price hours. Antweiler and Müsgens \cite{antweiler_2024} explore price-setting in a system with wind, solar, batteries, and hydrogen storage, allowing price-responsive demand. For the German system in 2050, four price levels result. VRES curtailment sets a price of zero in less than 10~\% of the time. The other price levels are set by hydrogen re-electrification with storage and batteries with different efficiencies, which provide firm capacity. They conclude EOMs remain functional with zero-marginal-cost generators if competition is ensured. Adachi et al. \cite{adachi_understanding_2024} investigate the market clearing prices, electricity demand and producer surplus in a simplified model on the transition to 100\% VRES. They find that average prices remain stable as long as price-responsive demand and storage are integrated into the energy market. \\
\newline

\subsection{Research Gap}

Research on market values and price formation in a decarbonised energy system frequently centres on particular technologies or oversimplified markets. While these studies offer valuable insights, they often lack a systematic approach to identifying the technologies that set market-clearing prices. Current analyses typically depend on heuristic methods, such as taking technologies in and out of the model, which are not universally applicable across different system configurations. To date, no general analytical framework exists to robustly determine price-setting technologies in complex, integrated energy systems. As a result, there is a lack of comprehensive understanding of how price-setting dynamics evolve throughout the transition from fossil-based to climate-neutral systems, particularly under conditions of high sector coupling and increased electrification. This represents a critical gap in the literature, limiting our ability to assess market behaviour and price formation in future decarbonised energy systems.

To address these research gaps, we introduce a new method of analysing the bidding behaviour of technologies that can be applied to a fully sector-coupled system. This enables us to reconstruct the full supply and demand curves ex-post, and thus determine which technologies are price-setting.
We can demonstrate how market clearing changes when transitioning from the current system to a climate-neutral energy system. This method enables us to identify the technologies that determine wholesale electricity prices in a sector-coupled, climate-neutral energy system. Consequently, we can demonstrate how and where prices emerge within the system and how this evolves over time. 

\section{Methods}
\label{sec:methods}

\subsection{Extracting price information}
\label{sec:price-infos}

In this section, we provide the mathematical representation of the optimisation problem and demonstrate how to extract the bidding curves ex-post using the stationarity equations for each component. This representation largely follows Brown et al. \cite{brown_price_2025}, which focuses on the power system, and adds complexity by incorporating cross-sectoral energy flows and conversion technologies. The energy system is modelled as a linear optimisation problem. By mapping the supply and demand bids to the dual variables (also known as \textit{shadow prices}) of the problem, we can determine the complete set of supply and demand technologies participating in the market clearing in every hour, including their bid prices and volumes. We can therefore reconstruct the entire merit order and derive the price-setting technology. To approximate electricity prices, we use the dual variable of the electricity balance constraint. However, this serves only as a proxy, since certain market effects are not captured by the bottom-up techno-economic model, as discussed in Section \ref{subsec:limitations}. \newline
\\
PyPSA represents the energy system as a hypergraph, with different energy carriers at the nodes, labelled by $i$, and edges labelled by $k$, representing energy conversion. Movement of energy carriers in time (storage) and space (networks) can be modelled as special cases of energy conversion. The spatial scope of the present study is limited to only one region, and consequently excludes networks and the associated costs. The modelled energy cariers include electricity, heat, (bio-)methane, coal, oil, hydrogen (+ derivatives) and biomass. Converters include heat pumps, electrolysers, etc. For clarity, the mathematical formulation in the main text omits temporal weighting factors, which corresponds to an hourly resolution of the problem. This simplification does not affect the underlying economic relationships or price formation mechanisms. A more general formulation including temporal weightings is provided in \ref{app:infos:methods:mathematical-problem-weightings}. \newline
The objective is to minimise the total annual system costs. These include investment costs as well as operational expenditures of energy supply, conversion and storage infrastructure. To express both as annual costs, we use an annuity factor that converts an asset's upfront investment into annual payments. As we use a myopic modelling approach with independent yearly optimisations, the time value of money across years is not considered. We obtain the annualised capital cost $c_*$ for investments at energy carrier $i$ in supply capacity $G_{i,r}\in\mathbb{R}^+$ of technology $r$, storage energy capacity ($E_{i,s}\in\mathbb{R}^+$) of technology $s$ and energy conversion capacities $F_k\in\mathbb{R}^+$. Furthermore, we apply the variable operating costs $o_*$ for supply dispatch $g_{i,r,t}\in\mathbb{R}^+$ and converter dispatch $f_{k,t}\in\mathbb{R}^+$ \cite{neumann_potential_2023}. 

\begin{align}
    \label{eq:objective}
    \min_{G,F,E,g,f} \quad & 
    \sum_{i,r} c_{i,r} G_{i,r} + \sum_{k} c_{k} F_k + \sum_{i,s} c_{i,s} E_{i,s} \notag \\
    &+ \sum_{i,r,t} o_{i,r} g_{i,r,t} +  \sum_{k,t} o_k f_{k,t}
\end{align}

For each energy carrier and each timestep, demand must be met by local generators, storage and energy flows from conversion units. 

\begin{align}
    \label{eq:nodal_balance}
    \sum_c d_{i,c,t} - \sum_r g_{i,r,t} + \sum_s (e_{i,s,t} - e_{i,s,t-1}) - \sum_k M_{i,k,t} f_{k,t} &= 0 \notag \\
    {} \quad \leftrightarrow \quad \lambda_{i,t} &\quad \forall i,t. 
\end{align}

where $d_{i,c,t}$ is the demand for energy carrier $i$ from consumer $c$ at timestep $t$, $e_{i,s,t}$ is the energy level of all storage units and expressed in carrier-specific units (e.g. MWh for electricity and heat, tonnes for CO$_2$). $M_{ikt}$ is the lossy incidence matrix of the network for the converters. It represents the connection (direction, efficiency and unit converison) between the energy carriers. It is a sparse matrix with non-zero values $-1$ when converter $k$ starts at node $i$ and $\eta_{i,k,t}$ if one of its terminal buses is node $i$. Here $\eta_{i,k,t}$ represents the conversion efficiency w.r.t. the starting node. $M_{ikt}$ can contain more than two non-zero entries if a converter has more than one input or output (e.g. gas combined heat and power (CHP) plant). The prices of all energy carriers $i$ correspond to the shadow price of the nodal balance constraint $\lambda_{i,t}$. These are also referred to as their marginal storage value (MSV). Consequently, the market clearing price is the shadow price of the electricity balance constraint $\lambda_{electricity,t}$.\newline
\\
The constraints can further be decomposed for each carrier $i$, storage medium $s$, and capacity constraints:\vspace{0.2cm}

\begin{minipage}{\columnwidth}
\begin{align}  
    - g_{i,r,t} &\leq 0  
    &&\leftrightarrow \ubar{\mu}^g_{i,r,t} \leq 0  \notag \\
    g_{i,r,t} - \bar{g}_{i,r,t} G_{i,r} &\leq 0  
    &&\leftrightarrow \bar{\mu}^g_{i,r,t} \leq 0 \notag \\
    - f_{k,t} &\leq 0  
    &&\leftrightarrow \ubar{\mu}^f_{i,k,t} \leq 0  \notag \\
    f_{k,t} - \bar{f}_{k,t}F_{k,t} &\leq 0  
    &&\leftrightarrow \bar{\mu}^f_{i,k,t} \leq 0 \notag \\
    - e_{i,s,t} &\leq 0  
    &&\leftrightarrow \ubar{\mu}^e_{i,s,t} \leq 0 \notag \\
    e_{i,s,t} - E_{i,s} &\leq 0  
    &&\leftrightarrow \bar{\mu}^e_{i,s,t} \leq 0 \notag \\
    \label{eq:constraints}
\end{align}
\vspace{0.2cm}
\end{minipage}

Here, $\bar{g}_{i,r,t}$ and $\bar{f}_{k,t}$ are the time and location-dependent availability factors, given per unit of nominal capacity. Must-run factors are not considered in the model. Additional constraints can be found in the Appendix \ref{app:infos:methods:add-constraints}\\

The atmosphere is represented as a storage unit with limited total capacity, where inflows correspond to emissions and the state of charge (SOC) to the cumulative stock of CO$_2$. The capacity represents the CO$_2$ emissions budget. The binding capacity constraint in the final period $T$ implies a shadow price $\bar{\mu}_{\text{CO$_2$},T}$, which corresponds to the CO$_2$ price. By the stationarity conditions of the KKT system, this value propagates backward in time, such that
\begin{align}
    \lambda_{\text{CO$_2$},t} = \bar{\mu}_{\text{CO$_2$},T} \quad \forall t.
\end{align}
Consequently, the marginal storage value of the CO$_2$ atmosphere directly represents the CO$_2$ price in \euro/tCO$_2$ in all timesteps.\\

The price formation can be derived from the stationarity equations of the Karush-Kuhn-Tucker (KKT) conditions. The derivation of the Lagrangian function is provided in \ref{app:infos:methods:lagrangian}. For the generation, we have

\begin{equation}
  0 =  \frac{\partial\mathcal{L}}{\partial g_{i,r,t}} = o_{i,r} - \lambda_{i,t} + \ubar{\mu}^g_{i,r,t} - \bar{\mu}^g_{i,r,t} \quad \forall i,r,t.
    \label{eq:generators-stationarity}
\end{equation}

The generator's variable costs $o_{i,r}$ and available capacity determine its bid into the market. If a generator $r$ is on the margin so that neither capacity constraint is binding, $\bar{\mu}^g_{i,r,t} = \ubar{\mu}^g_{i,r,t}  = 0$, the generator's variable cost is price-setting $\lambda_{i,t} = o_{i,r}$. If a generator's variable cost is below the price, it still bids with $o_{i,r}$ and $\bar{\mu}^g_{i,r,t}$ represents the inframarginal rent.  Conversely, if the generator's variable cost is above the price, it will not operate and $\ubar{\mu}^g_{i,r,t}$ will be negative. This is exactly how the steps of the conventional merit order curve are determined by the different variable costs $o_{i,r}$ with the widths determined by the capacity.\\

We now apply the same logic to the energy conversion technologies to determine how they bid. For each $f_{k,t}$, we have

\begin{equation}
  0 = \frac{\partial\mathcal{L}}{\partial f_{k,t}} = o_{k} - \sum_i M_{ikt} \lambda_{i,t}  + \ubar{\mu}^f_{i,k,t} - \bar{\mu}^f_{i,k,t} \quad \forall i,k,t.
  \label{eq:converter-stationarity}
\end{equation}

The ask/bid price for the converters, represented by the flows \( f_{k,t} \), is determined by the operating cost of the converter \( o_k \) and the shadow prices $\lambda_{i,t}$ of the other nodes connected, adjusted for the direction and efficiency as captured by \( M_{ikt} \). These represent opportunity costs for depleting the respective stored energy carriers, thus substituting the role of fuels for generators above. When no capacity constraint is binding, the carrier prices are related to the marginal cost by:
\begin{equation}
\sum_i M_{ikt} \lambda_{i,t} = o_k.
\end{equation}
This price reflects the cost of operating the converter and the efficiency of the energy carriers it connects as well as their prices.\\
\\

Let's assume the following specific case to derive the ask price \( \lambda_{j,t} \) for a gas turbine (converter) bidding into the electricity node $j$. The gas turbine $k$ connects two energy carriers $i$ (gas) and $j$ (electricity), with efficiency factor \( \eta_j \) for converting \( i \) to \( j \). \( M_{ikt} \) has two non-zero entries: \( M_{ikt} = -1 \) (indicating that \( i \) is an input to the converter) and \( M_{jkt} = \eta_j \) (indicating that \( j \) is an output of the converter). The converter flow \( f_{k,t} \) is not capacity constrained, so \( \ubar{\mu}^f_{k,t} = \bar{\mu}^f_{k,t} = 0 \). In this case, the relationship simplifies to:
\begin{equation}
- \lambda_{i,t} + \eta_j \lambda_{j,t} - o_k = 0.
\end{equation}
We can rearrange this to solve for \( \lambda_{j,t} \), the price that converter $k$ would ask for producing carrier \( j \):
\begin{equation}
\lambda_{j,t} = (\lambda_{i,t} + o_k) \eta_j^{-1} .
\label{eq:converter-ask-price}
\end{equation}

The ask price \( \lambda_{j,t} \) is determined by the price \( \lambda_{i,t} \) of the energy carrier \( i \) plus the operating costs per unit of carrier consumed \( o_k \) of the converter, adjusted for the efficiency \( \eta_j \). The price \( \lambda_{j,t} \) that a converter would ask for in a particular case depends on the conversion efficiency, the shadow price of the connected energy carrier, and the operating costs of the converter. When it is not price-setting and its ask price is lower than the clearing price, it will ask at full capacity, with $\bar{\mu}^f_{k,t} \leq 0$, which is the inframarginal rent. If its ask is higher than the price, it does not run and $\ubar{\mu}^f_{k,t} \leq 0$. Converters can be demand as well as supply technologies, depending on which end of the converter you take and the direction of the flow (e.g. electrolysis as hydrogen producer and electricity consumer). The bidding behaviour of a battery, as a capacity-constrained store, is modelled as a converter (see \ref{app:infos:methods:converter-storage-bidding}). It bids according to Equation \ref{eq:converter-ask-price} with $\lambda_{i,t}$ representing the value of one unit of stored energy and the discharging efficiency $\eta_j$. The marginal storage value $\lambda_{i,t}$ acts as the storage medium's "fuel cost" analogous to how coal or gas prices affect thermal generators. The derivation of bid and ask prices for a storage unit constrained only by its SOC is provided in \ref{app:infos:methods:storage-bidding}. For an explanation of the supply and demand volume bids, refer to \ref{app:infos:methods:volume-bids}.\newline
\\

\subsection{Obtaining the price setter from solved models}
\label{sec:price-setter}
Although the presented method is applicable to any energy carrier and market, this paper focuses on the electricity market. Assume that the bids from the electricity-related technologies are uniquely different. Then the price setter is the single technology that is running but not at full capacity and whose bid price equals the market clearing price (see \ref{sec:price-infos}). In conditions of perfect competition and high market liquidity, the market-clearing price is determined by the marginal technology. This is the technology with the highest bid or lowest ask that is still dispatched or consuming. However, the bids and therefore the marginal technology may not always be unique, e.g. due to model degeneracy and numerical inaccuracies. To determine the price setter in such cases, we first select a candidate set based on the following criteria:

\begin{enumerate}
  \item The difference of the reconstructed bid / ask price of a technology (see section \ref{sec:price-infos}) and the market clearing price cannot be larger than 0.01~\euro/MWh.
  \item The capacity utilisation of the technology must be smaller than 99\%. (\(g_{i,r,t} / \bar{g}_{i,r,t} G_{i,r} <  0.99\))
  \item The capacity utilisation of the technology must be larger than 1\% or the technology has to generate at least 10 MWh.
\end{enumerate}

If these criteria are met, the technology is added to the candidate set. We prioritise supply-side price setting over demand-side price setting. This approach aligns with conventional market logic, where the market-clearing price is typically determined by the marginal generation unit. In most electricity markets, demand is generally considered less elastic and less price-responsive, while supply bids tend to be more granular and price-sensitive. Although the role of demand-side participation is expected to grow in future markets, we adopt this convention for reasons of comparability with existing literature and methodological simplicity. Therefore, if a supply technology is in the candidate set, demand technologies are disregarded. If there is a unique supply technology in the candidate set, then it becomes the price setter. If there is no supply technology, but a unique demand technology, in the candidate set, then the demand technology is the price setter. The case with a single unique supply or demand technology occurs 28\% of the time across all years. In cases involving multiple candidates, we prioritise the least flexible technologies as the price setters. We measure flexibility by ordering the technologies according to the variance of their bids. A detailed description of the method and underlying assumptions can be found in Appendix \ref{app:infos:methods:price-setter-multiple}.

\subsection{Modelling of the German energy system}
\label{sec:modelling}
This section first introduces the general modelling framework (Section~\ref{sec:general-model}). All features described therein are subject to specific adaptations for this study, as outlined in Section~\ref{sec:study-specific-impl}.

\subsubsection{General model of the German energy system}
\label{sec:general-model}
We use a sector-coupled model of the Germany energy system: PyPSA-DE \cite{pypsa_de}. It is based on the open-source, multi-vector energy system model PyPSA-Eur \cite{horsch_pypsa-eur_2018}. This, in turn is built on the modelling framework Python for Power System Analysis (PyPSA) \cite{brown_pypsa_2018}.\newline
The model co-optimises the investment and operation of generation, storage, conversion and transmission infrastructures in a linear optimisation problem. It represents the energy system infrastructure for Germany and its neighbouring countries from 2020 to 2045. The model optimises supply, demand, storage, and transmission networks across a range of sectors including electricity, heating, transport, agriculture, and industry with myopic foresight. It is designed to minimise the total system costs by utilising a flexible number of regions and hourly resolution for full weather years.

PyPSA-DE provides a comprehensive representation of all major CO$_2$-emitting sectors. The CO$_2$ pricing is done indirectly via a budget as described in \ref{sec:price-infos}. Non-CO$_2$ greenhouse gas (GHG) emissions are taken into account by tightening the remaining CO$_2$ budget. This is necessary as PyPSA-DE only covers CO$_2$ emissions.\newline 
The heating sector is disaggregated into rural, urban decentral and urban central regions with options for district heating in urban central areas. The heating supply technologies include heat pumps, resistive heater, gas boilers, and CHP plants, which are optimised separately for decentralised use and central district heating. District heating networks can also be supplemented with waste heat from various conversion processes, including electrolysis, methanol production, ammonia synthesis, and Fischer-Tropsch fuel synthesis.

The transport sector is divided into three main segments: land transport (by energy source), shipping, and aviation. Main assumptions of the future development are a high degree of electrification in the land transport and usage of hydrogen and carbonaceous fuels in the aviation and shipping segments. The Aladin model provides German-specific data on e.g. the share of electrification in the transport sector \cite{plotz_modelling_2014}. Within the industrial sector, major sub-sectors such as steel, chemicals, non-metallic minerals, and non-ferrous metals are modelled separately to account for various fuel and process-switching scenarios. Fuels and processes as well as their transition paths are exogeneoulsy set. The Forecast model delivers data on production volumes for industry processes \cite{Fraunhofer2019forecast}. PyPSA-DE also includes different energy carriers and materials, with a focus on managing the carbon cycle through methods like carbon capture, BECCS, direct air capture, and using captured carbon for synthetic fuels. Renewable potentials and time series for wind and solar electricity generation are calculated using \textit{atlite}, taking into account land eligibility constraints such as nature reserves and distance criteria from settlements \cite{hofmann_atlite_2021}.\\
\\
PyPSA-DE encompasses a range of storage options to manage different forms of energy. Electricity can be stored in batteries (home, utility-scale, electric vehicles), pumped hydro  (PHS), hydrogen (produced via water electrolysis and stored in tanks or caverns), and synthetic energy carriers like methane and methanol. Hydrogen can be re-electrified using hydrogen turbines and fuel cells, while (synthetic) methane can be converted to electricity via gas turbines or CHP plants. Thermal energy storage is provided by large water pits for seasonal needs in district heating networks and short-term storage in individual home applications. Demand side management is also modelled, with battery electric vehicles (BEV) shifting loads during charging and electrolysis processes adjusting electricity use based on its price. More general details of PyPSA-DE can be found on the website of the Ariadne research project\footnote[1]{\href{https://ariadneprojekt.de/modell-dokumentation-pypsa/}{https://ariadneprojekt.de/modell-dokumentation-pypsa/}} and in Lindner et al. \cite{lindner_pypsa_2025}.\\
\\
A range of different technologies can be used for electricity generation and consumption, which affects the price setting. On the supply side, there are VRES, such as solar, solar-hsat (horizontal single-axis trackers), onshore and offshore wind, and run-of-river (ror). Conventional technologies include  open-cycle gas turbines (OCGT), closed-cycle gas turbines (CCGT), hard coal, lignite, and oil, which can be used as pure electricity generators or in cogeneration with heat. Gas turbines can also operate on upgraded biogas or synthetic methane. Storage technologies include PHS, reservoir hydroelectricity (hydro) and batteries. Biomass, methane and hydrogen can be combusted in turbines to generate electricity and/or heat. On the electricity demand side, the model includes inelastic loads for the residential and commercial sector, industry, and agriculture. More flexible consumers include air and ground heat pumps, resistive heaters, methanolisation, direct air capture (DAC), BEVs and electrolysis. Storage technologies include batteries and PHS. A critical exogeneous input for the price formation is the assumption on fuel costs and their development, which are summarised in Tab. \ref{tab:fuel-costs}.\newline

PyPSA-DE incorporates various Germany-specific policies. The CO$_2$ emission amounts are consistent with Germany`s GHG targets and can be seen in Tab. \ref{tab:co2_data}. The assumed sequestration limits in Germany for the different years can be taken from Tab. \ref{tab:co2_data}. Nuclear power plants are only included in 2020, following the political phase-out in 2023, and after 2038, coal-fired power plants are banned. From 2030 onwards, there is the possibility to build hydrogen-fuelled OCGT, CCGT and CHP. In 2035, all gas turbines built in 2030 are exogeneously retrofitted to use hydrogen. 
The model inlcudes lower capacity constraints on the buildout of renewable energy which are oriented towards the German goals until 2030.

\begin{table}[htbp]
\caption{CO$_2$ emissions and assumed CO$_2$ sequestration potential for Germany in Mt/a}
\begin{center}
\begin{tabular}{ccccccc}
\toprule
Year  & 2020 & 2025 & 2030 & 2035 & 2040 & 2045 \\
\midrule
CO$_2$ Emissions & 706 & 550 & 368 & 227 & 94 & -52 \\
CO$_2$ Sequestration & 0 & 0 & 10 & 20 & 40 & 80 \\
\bottomrule
\end{tabular}
\begin{tablenotes}
\item[a] CO$_2$ emissions for Germany based on \cite{ksg2021} and sequestration potential oriented to \cite{bmwk2024cms,luderer_energiewende_2025}.
\end{tablenotes}
\label{tab:co2_data}
\end{center}
\end{table}

\subsubsection{Study-specific modelling changes}
\label{sec:study-specific-impl}
This study reduces the complexity of PyPSA-DE by modelling Germany as a single copper-plated region with no connections to other countries. While this neglects cross-border effects on price formation, it simplifies the tracking of price-setting technologies. As the purpose of this study is to analyse price formation in integrated energy system transformations we focus on the domestic market and its dynamics, which we wanted to keep simple for the demonstration of this method. Other studies which incorporate cross-border effects, find that those effect are on the same magnitude and primarily smooth out the price levels through cross-border trade or different market structures in other countries \cite{bottger_wholesale_2021}. The model can import biomass in the range of 32 to 80 TWh/a. The amounts per year are based on recent modelling results for Germany \cite{luderer_energiewende_2025}. Furthermore, the system has the possibility to remove a maximum of 30 Mio. tonnes CO$_2$ from the German budget at a price of 300~\euro/tCO$_2$ in the year 2045. This is supposed to model carbon offsets outside of Germany. Load shedding is implemented at a price of 2000~\euro/MWh. It is assumed that there are no bottlenecks in the distribution grid. The weather data from 2019 is used, with a temporal resolution of 3 hours. Details on the optimised electricity balance time series and capacities of the resulting system can be found in the Appendix (see Figures~\ref{fig:elec-balance-all} and~\ref{fig:elec-capa}).

\section{Results}
\label{sec:results}

\subsection{General system changes from 2020 to 2045}
\label{sec:res:system-change}
To contextualize the development of electricity price setting, it is essential to examine the structural changes in the underlying energy system. Between 2020 and 2045, the installed capacities of solar and wind (onshore and offshore) generation expand significantly, from 54 GW and 62 GW to 570 GW and 298 GW, respectively. The land transport sector becomes fully electrified, resulting in an increase in electricity consumption from 16 TWh in 2020 to 206 TWh in 2045. Electrification of heating technologies contributes an additional 198 TWh of electricity demand by 2045. In the industrial sector, electrification leads to a rise in electricity consumption by approximately 170 TWh, corresponding to an increase of around 72\%. While only 8\% of electricity demand is modeled as flexible in 2020, this share increases to approximately 58\% by 2045, and is likely to be even higher if potential flexibilities in industry and household are considered. PtX technologies account for approximately 36\% of total electricity demand in 2045. Further general information on the electricity balance and capacities of the resulting system can be found in the Appendix (see Figures~\ref{fig:elec-balance-all} and~\ref{fig:elec-capa}).

\subsection{Price setting development}
\label{sec:res:price-setting}

\begin{figure}[htb]
    \centering
    \footnotesize
    (A) Development of price setting of electricity supply \\
    \includegraphics[width=\columnwidth]{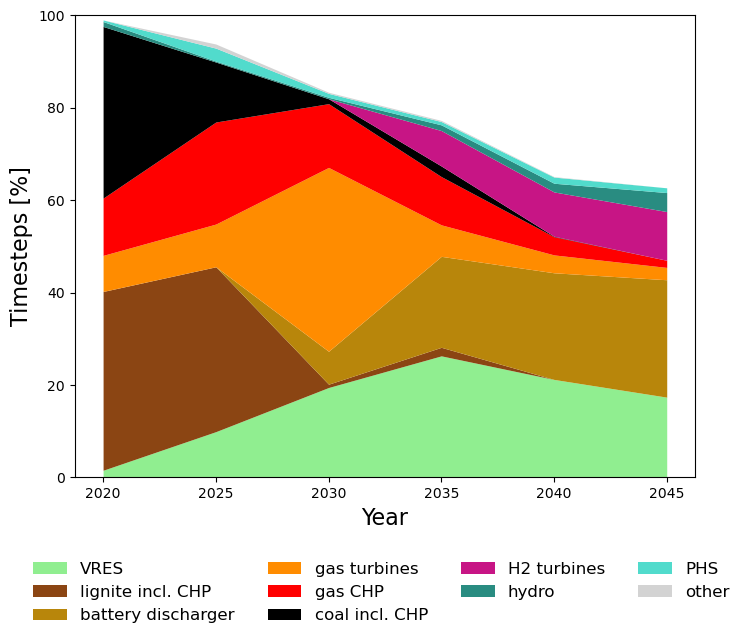} \\
    (B) Development of price setting of electricity demand \\
    \includegraphics[width=\columnwidth]{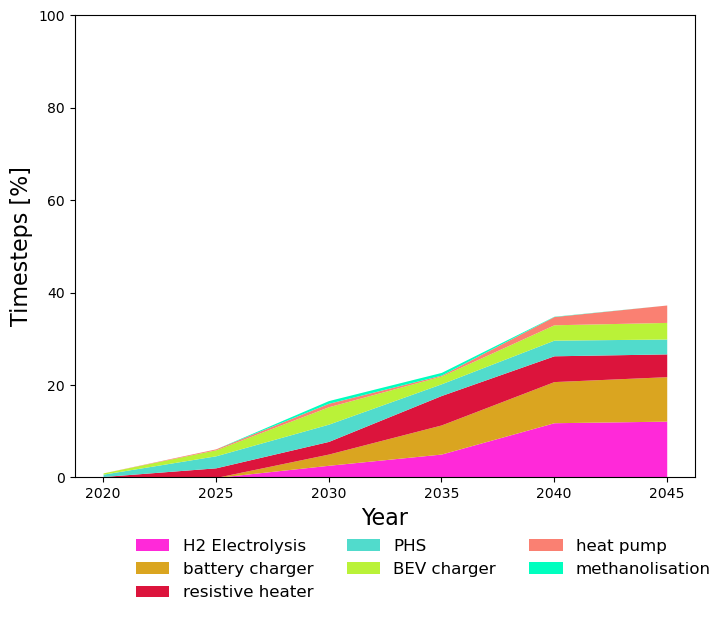} \\
    \caption{\textbf{Evolution of price-setting technologies.}
    The figure shows the number of timesteps at which a electricity supply (upper plot) or demand (lower plot) technology is price-setting in a specific year. Supply and demand add up to 100~\%.}
    \label{fig:price-setting-evo}
\end{figure}

Fig.~\ref{fig:price-setting-evo} summarises the evolution of price-setting technologies for the electricity market. Price setting develops from a system dominated by fossil fuel-based supply technologies to a system where prices are predominantly set by batteries, VRES and flexible demand technologies. Overall the percentage of supply technologies setting the price decreases from 97\% in 2020 to only 62\% in 2045. \newline

The price setting in 2020 and 2025 is dominated by coal power plants, which provide firm capacity and gas for price peaks. In 2030, the higher CO$_2$ price of 130~\euro/tCO$_2$ pushes the coal generation out of the market and gas turbines take over one part of their role in price setting. The other part is overtaken by battery discharge and an increasing curtailment of wind and solar. In 2035, the price is set mostly by the battery discharger and VRES. It is the year with the highest number of zero prices. The contribution of gas turbines is reduced and partly overtaken by the retrofitted hydrogen turbines. From 2035 on, the supply price setting only changes slightly. Gas turbines are almost entirely replaced by hydrogen turbines. Curtailment decreases as storage technologies and flexible demand enter the market, competing for low-price hours with bids that reflect their respective willingness to pay. The battery discharger consolidates its role as the most important price setter.\newline
\\
Until 2030, demand technologies play almost no role in the price setting. In 2030, especially the BEV chargers and PHS start to set prices. There is also a small contribution of battery charger and electrolysis, which become the most important price setters from 2035 on. PHS plays a moderate role throughout the whole development. The increased need for electricity balancing and therefore the higher degree of sector-coupling leads to a higher share of demand technologies setting the price. Demand technologies become more important in price setting with tighter CO$_2$ constraints. In 2020, only 3\% of the prices are set by demand, whereas in 2045 demand sets 38\% of the prices.\newline

\subsection{Market clearing in different situations and years}
\label{sec:res:market-clearing}

\begin{figure*}[p!]
    \centering
    \includegraphics[width=0.99\columnwidth]{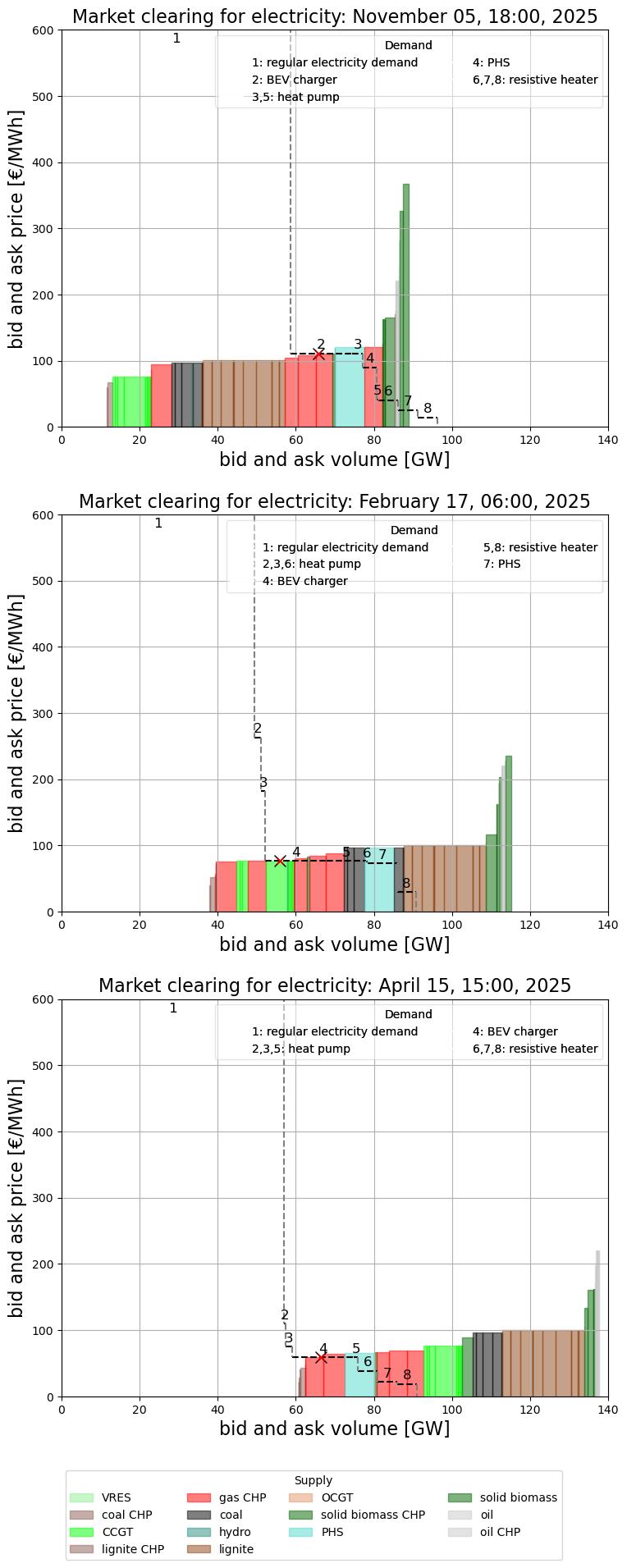} \qquad
    \includegraphics[width=0.99\columnwidth]{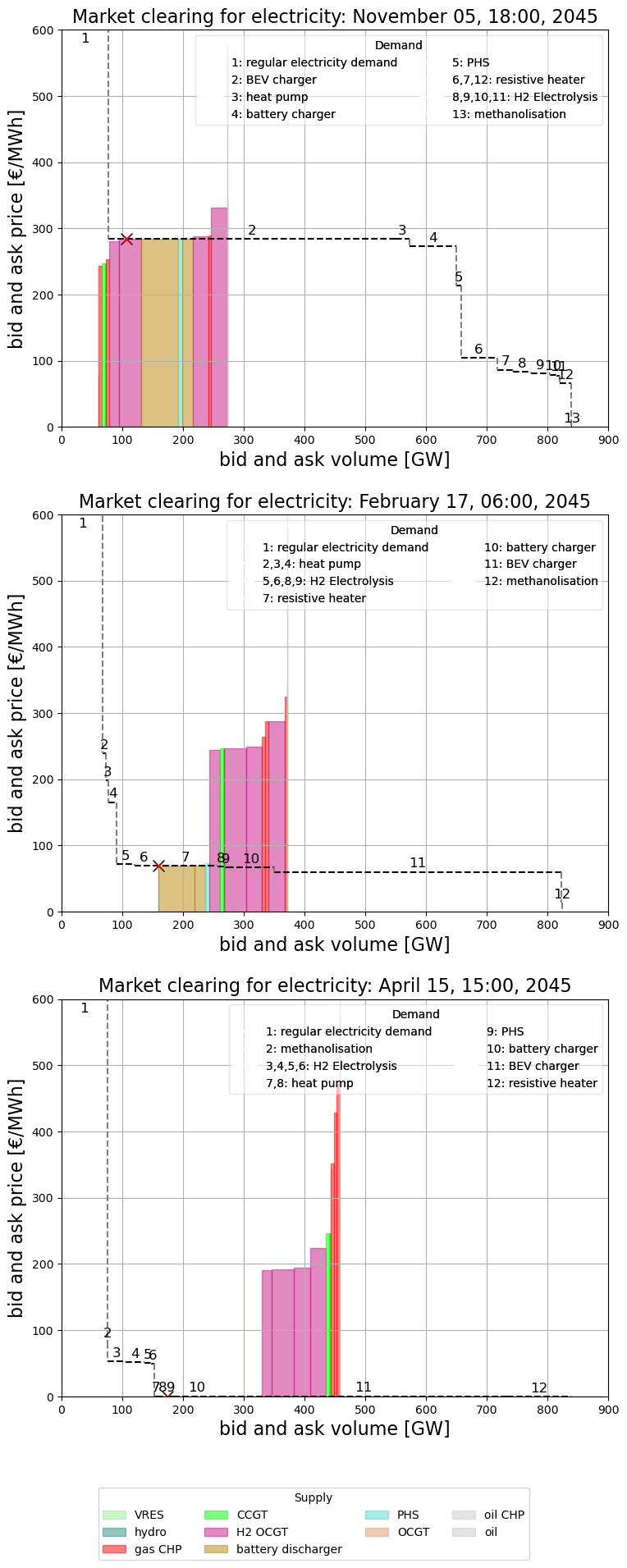}
    \caption{\textbf{Comparison of market clearing in representative situations of 2025 (left) and 2045 (right).}
    The figure shows the interaction of supply and demand for the electricity price formation in 2025 and 2045. The situations correspond to a price higher than the 0.7 quantile (top), between the 0.3 and 0.7 quantile (middle) and lower than the 0.3 quantile (bottom). Due to varying efficiencies and costs across different investment periods, technologies may appear more than once with different bid and ask prices.}
    \label{fig:merit-order}
\end{figure*}

In order to investigate the electricity price formation, we analyse the market clearing for different price situations and years. Fig.~\ref{fig:merit-order}  shows the interaction of supply and demand in a high, average and low price situation for the years 2025 and 2045. The high price is above the 0.7 quantile, the low price is below the 0.3 quantile, and the average price falls between these two thresholds. In 2025, the 0.3 and 0.7 quantiles correspond to 77~\euro/MWh and 100~\euro/MWh, respectively. In 2045, these thresholds shift to 50~\euro/MWh and 73~\euro/MWh. The most extreme values are reached in 2035 with 0.1~\euro/MWh and 121.3~\euro/MWh. The full set of price setting technologies separated by those quantiles can be found in the Appendix in Fig.~\ref{fig:price-setter-quantile}. \\

In general, high electricity prices in 2025 are set by lignite, gas CHP and PHS. Fig.~\ref{fig:merit-order} (left, upper) represents a winter day with high electricity demand and rather low renewable generation. On the demand side, there is the regular electricity demand for households, industry and agriculture, which is assumed to be perfectly inelastic. The remaining demand stems from heat technologies, PHS and BEV charger. On the supply side, wind, solar and ror are bidding with their marginal costs close to zero. Subsequently, fossil generators like coal, lignite and gas including CHPs make up most of the supply. Biomass and oil are the most expensive generators, which are not dispatched in this situation. The market is cleared by gas CHP at a price of 111~\euro/MWh. BEV charger and heat pumps bid exactly on the margin.\newline 

In the average price corridor of 77-100~\euro/MWh, coal, lignite, gas CHP and CCGT set the prices in 2025. In general, the merit order resembles the high-price scenario. In that year, no hydrogen turbines, synthetic methanation routes, or batteries are present in the system. As a result, CHPs, hydro, and PHS are the only supply technologies with variable bids. CHP units can vary their bids because heat revenues influence their bidding behaviour. The remaining part of the supply curve is constant owing to constant fuel costs. The increase of VRES bid volume to almost 40 GW shifts the supply curve to the right. Combined with the lower demand volume the intersection of supply and demand is at 77~\euro/MWh and the price is set by CCGT.\newline 

VRES curtailment as well as gas CHP, CCGT and coal CHP are the main price setters in the low price situations. As shown in Fig.~\ref{fig:merit-order} (left, lower), wind, solar and ror bid 60 GW into the market. Apart from those, only CHP technologies are running. Due to a moderately high price of 40~\euro/MWh at the heat bus, their electrcity bid is rather low. Gas CHP clears the market at 59~\euro/MWh.\\

The merit order in 2045 is fundamentally different from that in 2025. Firstly, there is much more VRES capacity with near-zero marginal costs, shifting the supply curve to the right. Secondly, the peak plants operating on fuel have much higher bid prices. Fossil fuels are expensive because of the price of the CO$_2$ certificates, while renewable fuels are expensive  to produce. This opens up a wider gap for more flexible technologies such as PHS and batteries, which do arbitrage between high and low price hours, to narrow the prices.\\
\\
In the high-price situations for clearing prices above 73~\euro/MWh, H$_2$ OCGT, battery discharger, methane CCGT, hydro and gas CHP set the prices. Fig.~\ref{fig:merit-order} (right, upper) shows an example situation in November. There is a relatively low bid volume from VRES and high ask prices for heating technologies. The bid prices of the merit order after the VRES bid start at 80~\euro/MWh with hydro and reach around 245~\euro/MWh with relatively low volumes. The merit order curve is much steeper than the one in 2025 and the volume of demand bids has increased from about 120 GW to more than 800 GW. Note that this represents the maximum theoretical consumption capacity—i.e., the case where all BEV chargers operate at full capacity simultaneously. While this scenario is unlikely, as a large share of the demand is shiftable, the maximum utilisation capacity in 2045 is 410 GW. Most heating technologies bid high into the market as they have to satisfy a non-elastic heat demand with limited storage options. Other technologies like PHS, battery charger, BEV chargers or electrolysis can bid more flexibly according to their marginal storage value (MSV). The price is set by the H$_2$ OCGT at 284~\euro/MWh.\newline

In an average price situation with prices ranging between 50 and 73~\euro/MWh, the market price is set by hydro, battery discharger and PHS from the supply side and mainly electrolysis and battery charger from the demand side. As shown in Fig.~\ref{fig:merit-order} (right, middle), a relatively high bid volume from variable VRES leads to price suppression. The only technology with a non-zero bid that is running is hydro. It also sets the price.\newline

For the low-price situations, which consist of prices below 50~\euro/MWh, solar, wind and battery are the main price setters. Overall the system in 2045 clears with prices below 1~\euro/MWh, which we classify as zero prices, for 25\% of the time. Note, that this does not necessarily represent a real-world situation as the model has no connection to other countries. This isolation leads to a need for more VRES generation and to more curtailment, resulting in a higher number of near-zero prices. With connection, the willingness to pay in neighbouring nodes would lead to higher clearing prices where the price is set by technologies outside of Germany. For the market clearing in Fig.~\ref{fig:merit-order} (right, lower), the VRES generation is very high and can satisfy all available demand volume. %
In this situation, there are only two technologies apart from the load that bid with non-zero prices: methanolisation and electrolysis. Both are running at full capacity. The electrolysis bid price is stabilised by a H$_2$ price of 78~\euro/MWh. The willingness to pay for heat is also close to zero because PtH technologies like heat pumps and resistive heaters are generating cheap heat and thermal energy storage makes it possible to take heat from the previous or following hour when electricity prices are low. Even though this has no effect on the market clearing price in this situation it shows that sector coupling, especially to energy carriers that are easier to store than electricity and heat can have a stabilising effect on the electricity (and heat) price. At prices such as these, basically, all electricity- and heat-related technologies are consuming at full capacity or until capacity is constrained by demand or storage availability.\\
\\

Comparing market clearing in 2025 and 2045 shows a fundamental change in the shape of the merit order. In 2025, the supply technologies are mostly based on fossil fuels and bid at fairly constant price levels. Renewable generation is not yet consistently high enough to meet the total demand and thus infrequently sets the market price. This leads to a merit order shape with a low bid volume of near-zero and a moderate increase thereafter, resulting in few near-zero prices. In 2045, the supply side is dominated by VRES. This leads to a drastic weather-dependent shift of the non-zero part of the merit order and to an increase in very high and very low prices. The transition from the VRES block to the non-zero block can be much steeper as the number of supply technologies is reduced and their average bids are increased by factors like the higher CO$_2$ price, high costs for synthetic fuels or scarcity cost on biomass. However, the transition slope from zero prices to the first price level is very variable. This is due to the ability of the battery chargers and PHS to store and dispatch the VRES generation according to their marginal storage values.\\
\\
On the demand side, the main difference is the increased volume in 2045 reaching up to more than 800 GW. In particular, there is more cross-sectoral demand of PtH technologies like heat pumps and resistive heaters, electrolysis and batteries. The demand for storable energy carriers such as hydrogen and methanol also increases. This has a stabilising effect on the electricity price as the bid price of those technologies is less dependent on the electricity price. For a more detailed analysis of the bidding behaviour of different technologies refer to \ref{app:infos:results:bid-ask}.

\subsection{Price duration}
\label{sec:res:price-duration}

\begin{figure}[htb]
    \includegraphics[width=\columnwidth]{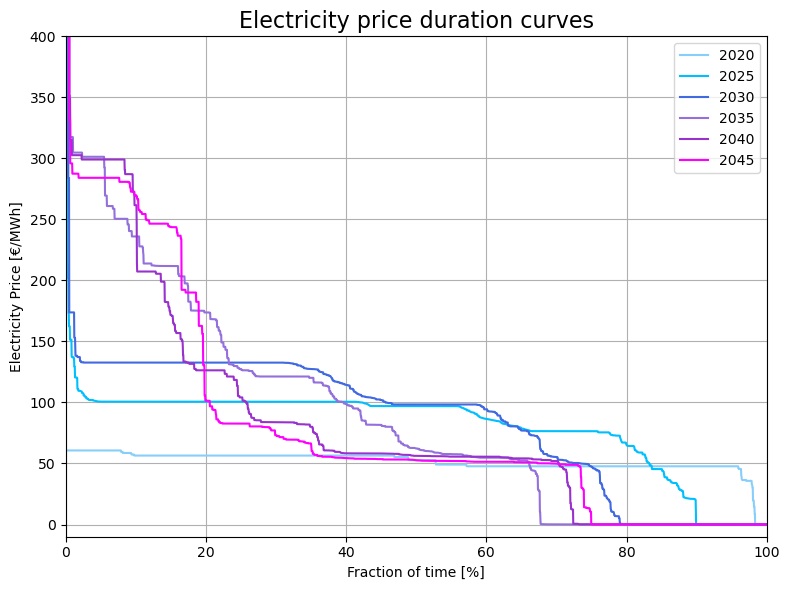}
    \caption{\textbf{Electricity price duration curves for different years representing different levels of CO$_2$ emissions.}
        The figure shows the development of the electricity price duration from the current system until a fully decarbonised system in 2045. The electricity prices above 400~\euro/MWh are shown in Fig. \ref{fig:elec-pdc-peak}.}
    \label{fig:elec-pdc}
\end{figure}

In order to gain insights into price formation and its evolution, we analyse the electricity price duration curves for different levels of CO$_2$ emissions represented by the years (Fig.~\ref{fig:elec-pdc}). To provide a clearer understanding of price setting, the price-setting technology is plotted over the price duration curves for all years in Fig.~\ref{fig:market-clearing-pdc-supply} (supply) and Fig.~\ref{fig:market-clearing-pdc-demand} (demand). These results should not be misinterpreted as price forecasts. There are several factors that can influence the price formation in the real world, such as the availability of interconnectors, the market design and the political framework which are not considered in detail in this model (see Section \ref{subsec:limitations}). The aim is rather to reveal general trends and patterns in the evolution of the price formation. 

To ensure that the price levels and distribution are roughly aligned with reality, we compare the prices from 2020 and 2025 with the wholesale electricity prices in Germany. The average day-ahead prices in Germany for 2020 and 2024 were 31~\euro/MWh and 80~\euro/MWh, respectively \cite{energycharts}. The average prices in the model for 2020 and 2025 are 52~\euro/MWh and 82~\euro/MWh. The number of hours with negative prices in Germany has been quite volatile in recent years. In 2020, there were 298 hours with negative prices; in 2022, this figure fell to 70 hours; and in 2024, it increased to 457 hours \cite{openenergytracker_negative_prices}. The model does not include any negative price hours because there are no subsidies or market premia for VRES and unit commitment is not considered. Prices below 1~\euro/MWh are classified as zero prices. According to the model results, this corresponds to 144 hours in 2020 and 882 hours in 2025. The model is expected to show a higher number of zero prices than in reality because there is no interconnection facilitating cross-border price convergence. In a study with the complete, interconnected model PyPSA-DE, the amount of zero prices did not exceed 10\% in any investment period \cite{luderer_energiewende_2025}. In general, this value is also very sensitive to the weather conditions and the regulatory framework. For this study we used weather data from 2019 consistently.\newline

A key finding is, that the range of price levels becomes wider with tighter CO$_2$ constraints. In earlier years, the electricity prices are set by generators whose bids are closer together. Fig.~\ref{fig:elec-pdc} shows periods of relatively constant price levels, particularly in the years 2020-2035. For 2040-2045, the electricity prices are distributed over a wider range.\\
\\
The reasons for this are manifold. The lower CO$_2$ price leads to bid prices of coal and gas-fired generators that are close to each other. With a CO$_2$ price of 28~\euro/tCO$_2$ in 2020, there are three major price levels  (see Fig.~\ref{fig:market-clearing-pdc-supply}). These are set by CCGT, lignite and hard coal. By comparison, the CO$_2$ price rises to 71~\euro/tCO$_2$ in 2025. This changes the order of the bid prices from CCGT, coal, and lignite due to their different CO$_2$ intensity. By 2030, the CO$_2$ price has risen to 131~\euro/tCO$_2$, widening the gap between fossil generators and pushing out coal-based generation.\\
\\
Additionally, the technology mix is changing. The difference between 2030 and 2035 cannot be explained by the price of CO$_2$ alone. The contribution of VRES generation and the need for balancing are constantly increasing. This results in a much smoother price curve over time, as storage technologies fill the gaps between less dominant price levels. Battery dischargers, in particular, set prices flexibly across a range of different price levels.\newline 

\begin{figure}[htb]
    \includegraphics[width=\columnwidth]{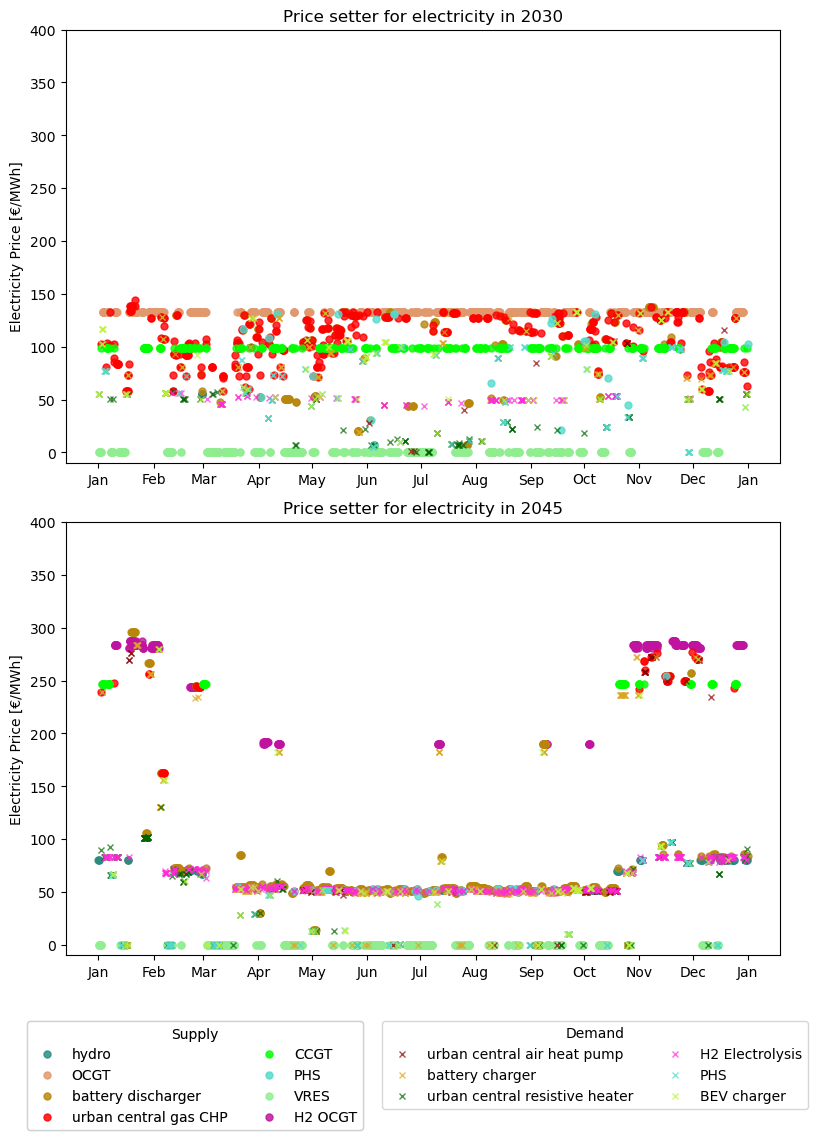}
    \caption{\textbf{Market clearing in 2030 vs. 2045 (temporal).}
        The figure shows the market clearing over time for the years 2030 and 2045. Indicated technologies correspond to the price setters.}
    \label{fig:market-clearing-temporal}
\end{figure}

For a better temporal understanding, Fig.~\ref{fig:market-clearing-temporal} shows the market clearing over time, comparing the years 2030 and 2045. For 2030 and the supply side, we can see the constant price levels set by OCGT and CCGT. Gas CHP especially drives winter prices and curtailments occur throughout the whole year. Especially in summer with substantial solar generation, electrolysis is setting the price at around 50~\euro/MWh.  The temporal market clearing for supply and demand and all years can be found in Fig.~\ref{fig:price-setting-temporal-supply} and Fig.~\ref{fig:price-setting-temporal-demand}.\\
\\

The combined effect of a higher CO$_2$ price and a changing technology mix can be seen in the market clearing of 2045 from Fig.~\ref{fig:market-clearing-temporal}. Coal and lignite no longer play a role due to the CO$_2$ price of 532~\euro/tCO$_2$. Peak prices occur mostly in winter set by H$_2$ OCGT, battery discharger, gas CHP and CCGT at around 280~\euro/MWh. In summer there are several price levels, which increase towards winter. The highest price level of around 200~\euro/MWh is set by H$_2$ OCGT, a mid level at 80~\euro/MWh is set by battery dischargers and hydro from the supply side and mainly electrolysis for the demand. The lowest level at around 50~\euro/MWh is mainly set by battery dischargers from the supply side. On the demand side, several technologies stabilise the prices at this level: electrolysis, battery charger and BEV charger.

The highest number of zero prices is reached on the transition. The price volatility and amount of zero prices are not monotonically increasing until the fully decarbonised system in 2045. The highest variance in prices and the most zero prices of 32\% are set in 2035 (see Fig.~\ref{fig:price-setter-quantile}). The reason for this is the fast buildout of VRES generation compared to the buildout of flexibility technologies and sector coupling. By 2045, the system exhibits sufficient flexibility throug PtH, electrolysis capacity, and storage, to absorb surplus electricity and thereby mitigate occurrences of zero prices. This emphasises the need for a fast buildout of not only the VRES generation but also the flexible demand linking to other sectors and storage. %

\subsection{Averaged supply and demand}
In order to analyse the evolution of price-setting supply structures, we calculate the average supply curve for each modelled year. For each timestep, we construct a stepwise supply curve. At each megawatt (MW), we calculate the average supply price across all timesteps, considering only points with sufficient data coverage. Specifically, the supply curve must contain data at that MW level for at least 5\% of the year. This yields a representative average supply curve for each year, showing structural shifts in the supply stack over time. The same procedure is applied to the demand side, where we calculate the average demand price for each MW. The resulting curves are shown in Fig.~\ref{fig:averaged-curves}. All the individual curves, including the average curves, can be found in the Appendix: Fig.~\ref{fig:supply-curves}, Fig.~\ref{fig:demand-curves}.\\\\

\begin{figure}[htb]
    \centering
    \footnotesize
    (A) Averaged supply curves \\
    \includegraphics[width=\columnwidth]{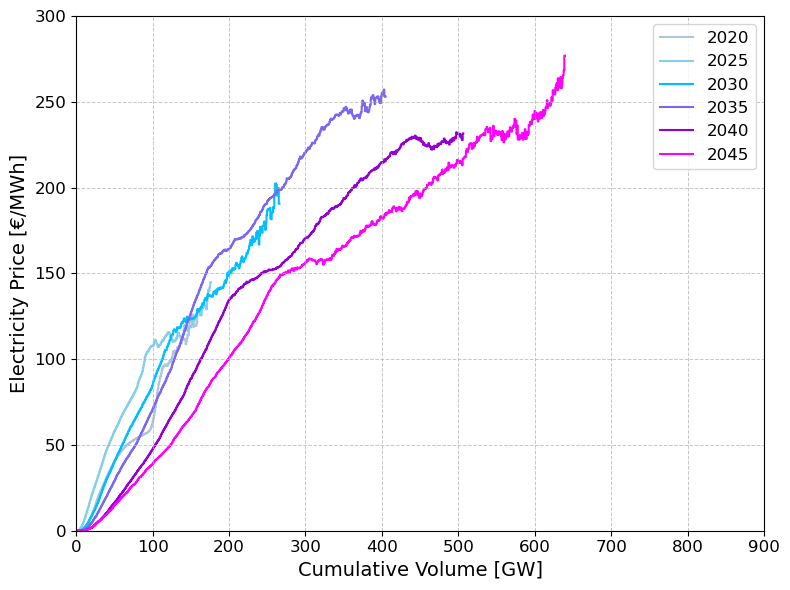} \\
    (B) Averaged demand curves \\
    \includegraphics[width=\columnwidth]{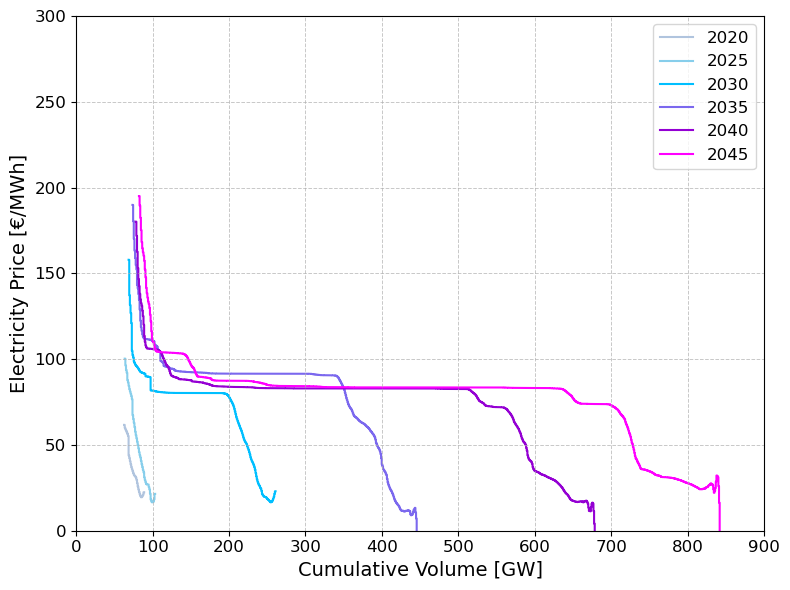} \\
    \caption{\textbf{Averaged supply and demand curves.}
    The figure shows the average supply and demand prices for every MW in the different years.}
    \label{fig:averaged-curves}
\end{figure}

The average supply curves exhibit a distinct flattening and rightward expansion over time. In 2020 and 2025, the total available supply volume is comparably low, with relatively steep price increases even at moderate cumulative volumes. As the system decarbonises, particularly from 2035 onwards, the supply curves become significantly flatter and extend beyond 500 GW of cumulative volume by 2045. This flatter start reflects increased installed capacity from near-zero marginal cost technologies like VRES in later years. However, the price increase is rather steep for the first 150-200 GW. At around 150~\euro/MWh the price increase is reduced until it transitions into a relatively flat peak price plateau.\\  

The average demand curves display the averaged willingness-to-pay of demand-side technologies. Over time, the curves span a wider price range and develop a much longer tail. This longer tail helps to absorb low-marginal-cost supply and avoid zero-price hours. Those in 2020 and 2025 are steep and narrow, indicating a limited range of demand-side technologies and a lack of flexibility. As the system decarbonises, the demand curves shift to the right and flatten out. From 2035 onward, the curves look very similar, except from the extending demand level at around 80~\euro/MWh. The curves also reach down to 0~\euro/MWh here. Until 100 GW, the willingness to pay is rather high, but falls rapidly. This is due to the demand for highly inelastic technologies such as PtH.

\section{Discussion}
\label{sec:discussion}

\subsection{Policy implications}
\label{sec:policy-implications}

The analysis of the electricity price formation reveals a significant transformation of the energy system from 2020 to 2045. Policy must take action to ensure that our carbon targets are met while price stability is maintained.

The results highlight a critical policy implication regarding the coordinated deployment of renewable generation and flexibility technologies during the energy transition. The period from 2030-2040 represents a particularly challenging transition period where rapid VRES buildout outpaces flexibility infrastructure development, resulting in the highest price volatility and zero-price hours. This finding underscores that energy system planning must ensure a balanced expansion of both variable renewable capacity and complementary flexibility resources to avoid inefficient market outcomes and maintain system stability during decarbonisation.\newline

The findings emphasise the importance of introducing dynamic pricing, which provides households and industries with access to real-time electricity prices (hourly or sub-hourly). This is essential for enabling demand-side flexibility, particularly from technologies such as electric vehicle charging, heat pumps, and industrial processes, which are vital for stabilising prices in a decarbonised energy system. Without these price signals reaching end users, the potential of these flexible resources remains untapped. Consequently, the electricity system may struggle to maintain the necessary balance between supply and demand for a stable market during the transition to renewable energy. A related aspect is investment security. Long-term investments, particularly in renewable energy and flexibility technologies, are typically guided by electricity price assumptions projected over 20 to 30 years. These assumptions critically influence project viability and investor confidence. Introducing more accurate, granular, and transparent price signals can reduce uncertainty and risk, thereby improving the quality of investment decisions.\newline

Sector-coupling technologies play a particularly crucial role in price stabilisation by providing demand-side flexibility that can absorb excess renewable generation. This effectively creates a price floor during periods of abundant renewable generation. The integration of PtX technologies and cross-sectoral demand from the heating and transport sectors thus serves as a critical buffer mechanism that prevents extreme price volatility while enabling efficient utilisation of VRES.\newline

Our results suggest that battery storage plays a pivotal role in balancing the future electricity markets and determining efficient prices. Additionally, the price setting of gas- and hydrogen-based technologies plays an important role for scarcity pricing. This highlights the importance of supportive policies that encourage the deployment of energy storage solutions at various timescales, from short-term battery storage for intraday balancing to long-term seasonal storage via power-to-gas technologies.\newline

\subsection{Relevance for other countries}
While the proposed method is broadly applicable, the specific findings derived for Germany may not directly translate to other national contexts. For instance, in countries like France, where the electricity mix is heavily dominated by nuclear power, market prices may remain influenced by nuclear generation due to its low marginal costs \cite{bottger_wholesale_2021}. Nations with abundant hydropower resources, such as Norway and Canada, may continue to rely on hydropower as the predominant marginal technology, owing to its inherent flexibility and substantial storage capabilities. In Germany, however, the potential for further hydropower expansion is highly constrained. In countries with limited dispatchable capacity and rapid renewable expansion, curtailment and the shift towards demand-side technologies in price setting may occur earlier.

\subsection{Expansion planning vs. operational planning}
\label{subsec:expansion-vs-operation}
The results shown were derived from a long-term (LT) model that co-optimises investment and operation. These largely align with the results of a short-term (ST) model that uses the the same already optimised capacities from the LT model as fixed and only performs dispatch optimisation. However, as discussed in Brown et al. \cite{brown_price_2025}, the results are not fully equivalent. From 2020 to 2040, the prices are nearly indistinguishable, while in 2045 the average price in the ST model is 12~\euro/MWh higher than in the LT model. The shapes of the price duration curves in 2045 are still very similar with a correlation of 0.99. The price-setting technologies do not differ in more than 3\% of the cases until 2045 and 10\% in 2045. The higher ST prices are mainly driven by the missing opportunity for the ST model to invest in new capacities \cite{brown_price_2025}. Convergence between the LT and ST can be forced by allowing a small elasticity for the regular electricity load, but this was found to be too computationally intensive, and the LT model is a better guide to the actual price setting. A comparison of prices between the LT and ST model and price-setting technologies of the ST model can be found in the Appendix: Fig. \ref{fig:pdc-st-lt} and Fig. \ref{fig:price-setting-evo-st}.

\subsection{Limitations} 
\label{subsec:limitations}
The study has several limitations.
It simplifies the energy system by modelling Germany as a single node without interconnections or grid constraints, allowing unrestricted energy transport within Germany and no exchange with other countries. Interconnection to other countries would lead to greater price heterogenity, smoother transitions, and fewer zero prices \cite{luderer_energiewende_2025,bottger_wholesale_2021}. The identification of price setters would still be possible with our method, but for the demonstration, we chose a simpler setup.\\ 
In the model we only consider backpressure CHP units, with a fixed heat-to-electricity ratio. More flexible extraction-condensing CHP units can have an impact on the price formation as shown in \cite{bottger_wholesale_2021}. 
Our approach uses dual information as a proxy for electricity prices. However, in reality electricity prices are influenced by many other factors that are not covered by the modelling approach. Specific examples include the absence of subsidies for VRES or maintenance periods for power infrastructure. The derived prices in this study stem from a centrally optimised model that optimises investment and operation across all technologies simultaneously. Consequently, they do not fully reflect real-world market dynamics, which are shaped by decentralised bidding behaviour, market structures, multi-market dependencies, order types and support instruments. While this approach captures fundamental supply and demand interactions, as well as technology-driven price signals, it ignores complex market mechanisms and the strategic behaviour of market participants. The model assumes perfect competition, rational behaviour, and the absence of market power. Due to perfect operational foresight within each of the investment periods, there are no operational uncertainties.\\
The price duration curves are sensitive to other factors like the used weather year, the modelling of the energy system components or investment and fuel costs.
The model does not incorporate unit commitment. This means start-up and shut-down costs are not included, nor are any constraints on minimum uptime, downtime or ramp-up rates. This is one reason why the model cannot be cleared at negative electricity prices. The absence of ramping constraints may also overestimate the operational flexibility of conventional power plants, affecting the system especially in earlier years.
In addition the demand is modelled as perfectly inelastic up to a value of lost load, whereas recent studies suggest that there exists a short-term price elasticity in Germany of about -0.05 mainly attributed to industrial electricity demand \cite{hirth_how_2024,arnold_functional_2023}. Although price elasticity was implemented in the model, it made computational feasibility challenging and was therefore left out the study. 

\section{Summary and conclusions}
We have introduced a new method for deriving bidding curves ex-post from sector-coupled energy system models. The method has allowed us to analyse electricity price formation in a model of the German energy system, revealing a significant transformation of the energy system from 2020 to 2045. In the early years, electricity prices are relatively stable, mainly driven by coal and gas-fired power plants. However, with rising CO$_2$ prices and the integration of VRES, 
storage and demand-side technologies begin to play an important role.\\
\newline

The price duration curves show an evolution from a state with several relatively stable price levels to a system with smoother transitions between less dominant price levels. The two main drivers are the CO$_2$ price and VRES penetration. The higher CO$_2$ price leads to a higher spread in the bid prices of the fossil fuel based price setters, resulting in a wider range of stable price levels. The higher VRES penetration leads to a stronger shift of the merit order curve, resulting in more situations where VRES are price-setting with near-zero marginal cost. This is reflected by the price-setting technologies. The system emerges from a state where predominantly coal and gas-fired power plants are price-setting to a system where VRES, battery and electrolysis determine the prices. Price volatility initially increases during the energy transition, peaking in 2035 with 32\% zero-price hours. By 2045, it stabilises as sufficient flexibility from storage, sector coupling, and electrolysis capacity emerges to absorb renewable energy surpluses.\\
\newline 

The merit order curve evolves from relatively flat price levels, which are dominated by fossil generators, to a much steeper and more weather-dependent structure with wider price ranges. VRES create a large near-zero price block followed by battery and peak prices set by hydrogen or gas turbines. This structural change fundamentally alters electricity market dynamics and pricing patterns. Cross-sectoral demand, particularly from PtH technologies and electrolysis along with demand for storable energy carriers such as hydrogen and methanol rises. The ask prices of technologies whose output is not electricity or heat but better storable carriers are less susceptible to a drop during hours of high VRES penetration. This leads to a stabilisation of the electricity and heat prices. As storable carriers like hydrogen and methanol become globally traded commodities, their prices will be shaped not only by local supply and demand but also by regional and international markets. These external price signals can propagate back into the electricity and heat sectors, creating new price dependencies that influence system dynamics, operational decisions, and investment strategies.\\
\newline

Along the decarbonisation pathway, the averaged supply curves flatten and shift outward, reflecting increased capacity from low marginal cost technologies like VRES, particularly after 2035. Early years show steep price increases at moderate volumes, while later years exhibit a high-volume plateau with reduced marginal price growth. On the demand side, curves broaden and flatten over time, indicating greater flexibility and a wider range of technologies. By 2045, demand extends to lower willingness-to-pay levels, with initial demand dominated by largely inelastic consumers such as PtH.\\
\newline

\subsection{Outlook}
The transition to a decarbonised energy system significantly alters market dynamics. Understanding these shifts is crucial for improving investment decision-making and for designing effective policies and market mechanisms that ensure a secure, affordable and sustainable energy supply. Future research should address the limitations of current models, such as representing multiple regions, identifying price-setting technologies across market zones and interconnectors, integrating elastic demand and evaluating alternative market designs. Additionally, the analysis should encompass countries with different technology mixes. Cross-border trading has a significant impact and will eventually need to be analysed with an extended spatial resolution and scope. One immediate area for future research is investigating the timing between VRES and flexibility buildout. Future work should also examine the role of battery storage in distribution grids, including peak shaving, remuneration schemes, and interactions with short-term markets. 
\section*{Acknowledgements}
J.G. and M.L. gratefully acknowledge funding from the Kopernikus-Ariadne project by the German Federal Ministry of Research, Technology and Space (Bundesministerium für Forschung, Technologie und Raumfahrt, BMFTR), grant number 03SFK5R0-2.

\section*{Author contributions}

\textbf{J.G.}:
Conceptualisation --
Data curation --
Formal Analysis --
Investigation --
Methodology --
Software --
Validation --
Visualisation --
Writing - original draft --
\textbf{F.N.}:
Investigation --
Methodology --
Software --
Supervision --
Validation --
Writing - review \& editing
\textbf{M.L.}:
Investigation --
Methodology --
Software --
Supervision --
Validation --
Writing - review \& editing
\textbf{P.H.}:
Validation --
Writing - review \& editing
\textbf{T.B.}:
Conceptualisation --
Formal Analysis --
Funding acquisition --
Investigation --
Methodology --
Project administration --
Supervision --
Validation --
Writing - review \& editing

\section*{Declaration of interests}

The authors declare no competing interests.

\section*{Declaration of generative AI and AI-assisted technologies in the writing
process}

During the preparation of this work the author(s) used ChatGPT in order
to improve wording. After using this tool/service, the author(s) reviewed and edited
the content as needed and take(s) full responsibility for the content of the published
article.

\section*{Data and code availability}
A dataset of the model results will be made available on \url{zenodo} after peer-review.
The code to reproduce the experiments is available at \url{https://github.com/JulianGeis/pypsa-de-pricing/tree/one-node-pricing}.

\renewcommand{\ttdefault}{\sfdefault}

\onecolumn
\section{Appendix}
\begin{appendices}
\section{Additional Information}
\label{app:infos}

\subsection{Methods}
\label{app:infos:methods}

\subsubsection{Mathematical problem with temporal weightings}
\label{app:infos:methods:mathematical-problem-weightings}

Including temporal weighting factors $w_{t}$, the objective function of the problem is given by:

\begin{align}
    \label{eq:objective_w}
    \min_{G,F,E,g,f} \quad & 
    \sum_{i,r} c_{i,r} G_{i,r} + \sum_{k} c_{k} F_k + \sum_{i,s} c_{i,s} E_{i,s} \notag \\
    &+ \sum_{t} w_{t} \left( \sum_{i,r} o_{i,r} g_{i,r,t} +  \sum_{k} o_k f_{k,t} \right)
\end{align}

Representative time snapshots $t$ are assigned weights $w_{t}$ such that their total duration sums to 8,760 hours. This ensures consistency between quantities expressed in MW, such as generator dispatch, and those expressed in MWh, such as energy storage. Since this is primarily an implementation detail and does not contribute to the conceptual understanding, it is omitted from the main text. The nodal balance constraint changes accordingly:

\begin{align}
    \label{eq:nodal_balance_w}
    \sum_c d_{i,c,t} - \sum_r g_{i,r,t} + \sum_s w_{t}^{-1}(e_{i,s,t} - e_{i,s,t-1}) - \sum_k M_{i,k,t} f_{k,t} &= 0 \notag \\
    {} \quad \leftrightarrow \quad w_{t} \lambda_{i,t} &\quad \forall i,t. 
\end{align}

where $d_{i,c,t}$ is the demand for energy carrier $i$ from consumer $c$ at timestep $t$, $e_{i,s,t}$ is the energy level of all storage units and expressed in carrier-specific units (e.g. MWh for electricity and heat, tonnes for CO$_2$). $M_{ikt}$ is the lossy incidence matrix of the network for the converters. It represents the connection (direction, efficiency and unit converison) between the energy carriers. It is a sparse matrix with non-zero values $-1$ when converter $k$ starts at node $i$ and $\eta_{i,k,t}$ if one of its terminal buses is node $i$. Here $\eta_{i,k,t}$ represents the conversion efficiency w.r.t. the starting node. $M_{ikt}$ can contain more than two non-zero entries if a converter has more than one input or output (e.g. gas combined heat and power (CHP) plant). The prices of all energy carriers $i$ correspond to the shadow price of the nodal balance constraint $\lambda_{i,t}$. Consequently, the market clearing price is the shadow price of the electricity balance constraint $\lambda_{electricity,t}$.\newline
\\
The constraints can further be decomposed for each carrier $i$, storage medium $s$, and capacity constraints:\vspace{0.2cm}

\begin{minipage}{\columnwidth}
\begin{align}  
    - g_{i,r,t} &\leq 0  
    &&\leftrightarrow w_{t} \ubar{\mu}^g_{i,r,t} \leq 0  \notag \\
    g_{i,r,t} - \bar{g}_{i,r,t} G_{i,r} &\leq 0  
    &&\leftrightarrow w_{t} \bar{\mu}^g_{i,r,t} \leq 0 \notag \\
    - f_{k,t} &\leq 0  
    &&\leftrightarrow w_{t} \ubar{\mu}^f_{i,k,t} \leq 0  \notag \\
    f_{k,t} - \bar{f}_{k,t}F_{k,t} &\leq 0  
    &&\leftrightarrow w_{t} \bar{\mu}^f_{i,k,t} \leq 0 \notag \\
    - e_{i,s,t} &\leq 0  
    &&\leftrightarrow \ubar{\mu}^e_{i,s,t} \leq 0 \notag \\
    e_{i,s,t} - E_{i,s} &\leq 0  
    &&\leftrightarrow \bar{\mu}^e_{i,s,t} \leq 0 \notag \\
    \label{eq:constraints}
\end{align}
\vspace{0.2cm}
\end{minipage}

Note that while the constraints themselves are not affected by the weighting of time snapshots $w_t$, the corresponding dual variables (e.g., $\lambda_{i,t}$, $\overline{\mu}^g_{i,r,t}$) must be interpreted as scaled by $w_t$, due to the objective function's time-weighted formulation. This ensures consistent units and correct economic interpretation of shadow prices (\euro/MWh, etc.).

\subsubsection{Additional constraints}
\label{app:infos:methods:add-constraints}

The capacities of generation, storage and conversion infrastructure are constrained from above by their installable potentials and from below by any existing components:

\begin{align}
    \underline{G}_{i,r} &\leq G_{i,r} \leq \overline{G}_{i,r} &&\forall i, r \notag \\
    \underline{E}_{i,s} &\leq E_{i,s} \leq \overline{E}_{i,s} &&\forall i, s \notag \\
    \underline{F}_{k} &\leq F_{k} \leq \overline{F}_{k} &&\forall k 
    \label{eq:capacity_constraints}
\end{align}

Some generators also additionally have a generation volume limit:
\begin{align}
\sum_t g_{i,r,t} \leq \overline{g}_{i,r} &&\leftrightarrow \mu^V_{i,r} &&\forall i, r
\end{align}

Here, $\mu^V_{i,r}$ denotes the dual variable (shadow price) of the generation volume limit. A non-zero value of $\mu^V_{i,r}$ indicates scarcity in the available generation volume, which can, in turn, affect the bids these generators submit in the market model.

The energy levels of the storage $e_{i,s,t}$ are either cyclic or given an initial state of charge (SOC). $\mathcal{T}$ represents the set of all time snapshots.

\begin{align}
    e_{i,s,0} - e_{i,s,|\mathcal{T}|} &= 0 
    &&\leftrightarrow \lambda^{\text{cyc}}_{i,s} \quad \forall i,s \notag \\
     e_{i,s,0} - e_{i,s, initial} &= 0 
    &&\leftrightarrow \lambda^{\text{init}}_{i,s} \quad \forall i,s   
    \label{eq:storage_constraints}
\end{align}

\subsubsection{Lagrangian}
\label{app:infos:methods:lagrangian}

The Lagrangian function for the problem is given by:

\begin{equation}
    \mathcal{L}(x, \lambda, \mu) = f(x) - \sum_{i} \lambda_i [g_i(x) - c_{i}] - \sum_{j} \mu_j [h_j(x) - d_{j}]
    \label{eq:lagrangian-general}
\end{equation}
where \( f(x) \) is the objective function, \( g_i(x) \) are the equality constraints, and \( h_j(x) \) are the inequality constraints. The variables \( \lambda_i \) and \( \mu_j \) are the Lagrange multipliers associated with the equality and inequality constraints, respectively.

For the generators $g_{i,r,t}$, the Lagrangian is given by:
\begin{equation}
    \mathcal{L}(g_{i,r,t}, \lambda, \mu) = \sum_{i,r,t} o_{i,r} g_{i,r,t} - \lambda_{i,t} g_{i,r,t} - (-g_{i,r,t}) \ubar{\mu}^{g}_{i,r,t} - (g_{i,r,t} - G_{i,r,t} G_{i,r}) \bar{\mu}^{g}_{i,r,t}
    \label{eq:lagrangian-generators}    
\end{equation}

For the conversion units $f_{k,t}$, the Lagrangian is:
\begin{equation}
    \mathcal{L}(f_{k,t}, \lambda, \mu) = \sum_{k,t} o_k f_{k,t} - \sum_i M_{ikt} \lambda_{i,t} f_{k,t} - (-f_{k,t}) \ubar{\mu}^{f}_{k,t} - (f_{k,t} - F_k) \bar{\mu}^{f}_{k,t}
    \label{eq:lagrangian-converters}
\end{equation}

\subsubsection{Bidding of storage represented as converter}
\label{app:infos:methods:converter-storage-bidding}
The dispatch of a capacity-constrained storage unit can be represented as a special case of a conversion technology, wherein stored energy is converted into usable energy. In PyPSA-DE, this modeling approach is applied to technologies such as batteries, PHS and reservoir-based hydro. While the following discussion focuses on batteries as an illustrative example, the underlying formulation is applicable to any storage technology with capacity constraints. In this framework, the storage medium is treated as a distinct energy carrier. The battery discharger connects two energy carriers: the stored energy in the battery (carrier $i$) and electricity (carrier $j$). The conversion is characterised by the discharge efficiency $\eta_j$ and operating costs $o_k$.
Following the general conversion unit formulation from Equation \ref{eq:converter-stationarity} without the temporal weighting factor, the stationarity condition for the battery discharger dispatch $f_{k,t}$ is:
\begin{equation}
0 = \frac{\partial\mathcal{L}}{\partial f_{k,t}} = o_k + \lambda_{i,t} - \eta_j \lambda_{j,t} + \ubar{\mu}^f_{k,t} - \bar{\mu}^f_{k,t} \quad \forall k,t,
\end{equation}
where $\lambda_{i,t}$ represents the marginal storage value (MSV) of the stored energy and $\lambda_{j,t}$ is the electricity price. When the battery discharger is not capacity constrained, so that $\ubar{\mu}^f_{k,t} = \bar{\mu}^f_{k,t} = 0$, the relationship simplifies to:
\begin{equation}
\lambda_{j,t} = (\lambda_{i,t} + o_k) \eta_j^{-1}.
\end{equation}
This equation reveals that the battery discharger bids into the electricity market with an effective bid price of $(\lambda_{i,t} + o_k) \eta_j^{-1}$. The marginal storage value $\lambda_{i,t}$ acts as the storage medium's "fuel cost," analogous to how coal or gas prices affect thermal generators. The battery will discharge at full capacity (and $\bar{\mu}^f_{k,t} \geq 0$) when the electricity price exceeds this bid price, and if the battery is price-setting, the electricity price equals the battery's effective bid.\newline 
Analogously, the battery charger converts electricity (carrier $j$) into stored energy (carrier $i$), characterised by the charging efficiency $\eta_l$ and operating costs $o_k$. When not constrained, the stationarity condition yields an effective ask price of $(\lambda_{i,t} - o_k) \eta_l$, at which the battery is willing to purchase electricity from the market. The battery charges when the market price falls below this ask price, and if it is the marginal technology, it sets the electricity price accordingly. This formulation demonstrates how energy storage participates in price formation through its marginal storage value, treating stored energy as a fuel with its own price ($\lambda_{i,t}$). This price reflects the temporal arbitrage value of the storage resource. The MSV captures the opportunity cost of using stored energy versus saving it for higher-value periods.

\subsubsection{Storage bidding}
\label{app:infos:methods:storage-bidding}
The previous analysis examined how storage influences price formation through its charging and discharging via converters. However, storage can also directly set prices through the marginal value of the stored energy itself. This is modelled as a storage unit that is only constrained by its state of charge (SOC) and not by charging and discharging capacity. Although this mechanism is relevant in general, it does not apply in the electricity market modeled here, as all electricity storage units are connected via converters. Nonetheless, we include this analysis for completeness and to illustrate the broader principles of how storage can influence price formation. This example shows the case for a storage with cyclic storage levels (see Equation \ref{eq:storage_constraints}). We first derive the Lagrangian with respect to the storage energy level variables $e_{i,s,t}$ at different points in time.\\

The Lagrangian for the storage energy levels is given by:
\begin{equation}
    \mathcal{L} = \sum_t \lambda_{i,t} \left( e_{i,s,t} - e_{i,s,t-1} \right) - \lambda^{\text{cyc}}_{i,s} \left( e_{i,s,0} - e_{i,s,|\mathcal{T}|} \right) - \sum_t \underline{\mu}^e_{i,s,t} \left( - e_{i,s,t} \right) - \sum_t \overline{\mu}^e_{i,s,t} \left( e_{i,s,t} - E_{i,s} \right)
    \label{eq:storage-lagrangian}
\end{equation}

Initial storage level $e_{i,s,0}$:
\begin{equation}
    \frac{\partial\mathcal{L}}{\partial e_{i,s,0}} = \lambda_{i,0} - \lambda_{i,1} - \lambda^{\text{cyc}}_{i,s} + \underline{\mu}^e_{i,s,0} - \overline{\mu}^e_{i,s,0} = 0
\end{equation}

This condition reflects the impact of the initial SOC on the first-period shadow price $\lambda_{i,0}$, the MSV at time $t = 1$ ($\lambda_{i,1}$), the cyclicity constraint $\lambda^{\text{cyc}}_{i,s}$, and the storage capacity constraints at time $t = 0$.\\

Intermediate storage levels $e_{i,s,t}$ ($1 \leq t \leq |\mathcal{T}| - 1$):
\begin{equation}
    \frac{\partial\mathcal{L}}{\partial e_{i,s,t}} = \lambda_{i,t} - \lambda_{i,t+1} + \underline{\mu}^e_{i,s,t} - \overline{\mu}^e_{i,s,t} = 0
\end{equation}

This condition links the marginal value of stored energy across consecutive time steps via the shadow prices $\lambda_{i,t}$ and $\lambda_{i,t+1}$, and incorporates the active lower and upper capacity constraints on storage.\\

Final storage level $e_{i,s,|\mathcal{T}|}$:
\begin{equation}
    \frac{\partial\mathcal{L}}{\partial e_{i,s,|\mathcal{T}|}} = \lambda_{i,|\mathcal{T}|} - \lambda^{\text{cyc}}_{i,s} + \underline{\mu}^e_{i,s,|\mathcal{T}|} - \overline{\mu}^e_{i,s,|\mathcal{T}|} = 0
\end{equation}
This final condition captures the influence of the marginal price at the end of the time horizon and includes the effect of the cyclic storage constraint and capacity constraints.
\subsection*{Price setting:}

Storage energy levels can directly determine prices. If they do their associated capacity constraints are not binding, i.e., the dual variables $\underline{\mu}^e_{i,s,t} = \overline{\mu}^e_{i,s,t} = 0$. In this case, the Lagrangian stationarity conditions simplify and allow for a direct relationship between storage levels and locational marginal prices. 

At the initial time step, the stationarity condition reduces to $\lambda_{i,0} - \lambda_{i,1} - \lambda^{\text{cyc}}_{i,s} = 0$. Solving for the price yields:
\begin{equation}
\lambda_{i,0} = \lambda_{i,1} + \lambda^{\text{cyc}}_{i,s}
\end{equation}
Thus, the price in the first period is associated to the MSV of the consecutive time step and the multiplier of the cyclic storage constraint. 
Within intermediate storage levels, the corresponding stationarity condition becomes:
\begin{equation}
\lambda_{i,t} = \lambda_{i,t+1}
\end{equation}
Hence, storage levels enforce a flat price profile across consecutive time periods whenever they are price setting. For the final storage level, if the capacity constraints are not binding, the stationarity condition simplifies to:
\begin{equation}
\lambda_{i,|\mathcal{T}|} = \lambda^{\text{cyc}}_{i,s}
\end{equation}
Thus, similar to the initial time step, the price in the final period is also determined by the cyclic constraint multiplier. The marginal storage value $\lambda_{i,t}$ represents the shadow price of stored energy. When storage energy levels directly set the price, several important properties follow. In both the initial and final periods, the prices are directly linked to the cyclic constraint multiplier $\lambda^{\text{cyc}}_{i,s}$, which ensures that the storage level at the end of the optimisation period equals the initial level. For intermediate time steps, when storage is the marginal price setting technology, it imposes a temporal arbitrage condition that equalizes the opportunity cost of storing versus using energy across different time periods. It is worth noting that the model assumes perfect foresight over the planning horizon, meaning that storage dispatch decisions and resulting shadow prices are based on full knowledge of future system states. In reality, forecasting uncertainty and the unknown value of stored energy at the end of the horizon complicate the valuation of storage.

\subsubsection{Supply and demand volume bids}
\label{app:infos:methods:volume-bids}

The volume bid of a technology represents its maximum available capacity. For generators, this is limited by their installed capacity and the time-dependent availability factor $\bar{g}_{i,r,t}$. The supply volume bid $S$ for generator $r$ at energy carrier $i$ is restriced as follows:

\begin{equation}
S_{g_{i,r,t}} \leq \bar{g}_{i,r,t} \cdot G_{i,r} \quad \forall i,r,t.
\end{equation}

Conversion technologies bid with their available dispatch capacity ($\bar{f}_{k,t}F_{k,t}$), multiplied by the efficiency of the corresponding output carrier ($\eta_{j,k,t}$). The volume demand of converters is determined by their withdrawal capacity for the respective energy carrier. For converter $k$ operating between energy carriers, the supply volume bid for output carrier $j$ and demand volume bid $D$ for input carrier $i$ are:

\begin{align}
S_{f_{j,k,t}} \leq \bar{f}_{k,t}F_{k,t} \cdot \eta_{j,k,t} \quad \forall j,k,t \\
D_{f_{i,k,t}} \leq \bar{f}_{k,t}F_{k,t} \quad \forall i,k,t.
\end{align}

In the model, we distinguish between two types of storage technologies. General storage units, such as gas and hydrogen storage, are only constrained by their state of charge (SOC). Their volume bid is limited by the current SOC and the discharge efficiency $\eta^{discharge}_{i,s}$, while their volume demand is constrained by the remaining space in the storage unit and the charge efficiency $\eta^{charge}_{i,s}$:

\begin{align}
S_{e_{i,s,t}} \leq e_{i,s,t-1} \cdot \eta^{discharge}_{i,s} \quad \forall i,s,t \\
D_{e_{i,s,t}} \leq (E_{i,s} - e_{i,s,t-1}) \cdot (\eta^{charge}_{i,s})^{-1} \quad \forall i,s,t.
\end{align}

In contrast, capacity-constrained storage units such as batteries, PHS and hydro are limited both by the SOC and their charging and discharging capacities implied by the converters. Their volume bid is defined as the minimum of the available dispatch capacity and the SOC, both adjusted by the discharge efficiency. The volume demand is the minimum of the available charge capacity and the remaining storage space. Here converter k is associated with storage unit s. 

\begin{align}
S_{e_{i,s,t}} \leq \min(\bar{f}_{k,t}F_{k,t}, e_{i,s,t-1} \cdot \eta^{discharge}_{i,s}) \quad \forall k,i,s,t \\
D_{e_{i,s,t}} \leq \min(\bar{f}_{k,t}F_{k,t}, (E_{i,s} - e_{i,s,t-1}) \cdot (\eta^{charge}_{i,s})^{-1}) \quad \forall k,i,s,t.
\end{align}

For PHS and hydro, the SOC additionally depends on natural inflows, spillage and standing losses.

\subsubsection{Obtaining the price setter for cases with more than one candidate}
\label{app:infos:methods:price-setter-multiple}

In cases involving multiple candidates, we designate the technology with the least ability to alter its prices as the price setter. Technologies that can only bid at one price level include generators with constant marginal costs and conversion units with fixed input node prices. These include VRES, nuclear, coal, lignite and biomass technologies. Oil and gas units have a constant fossil fuel input price in the model. However, in addition to these constant fossil-based prices, there are also synthetic and biogenic production routes, which can introduce variability in price setting. CHP units, PtX technologies, and storage technologies exhibit a higher variability in their bid prices due to the observed variability in input prices and secondary effects.\newline
In equilibrium, the prices at the different nodes can be equalised in such a way that they differ only by the conversion efficiency of certain technologies. This often results in the more adaptable/flexible technologies bidding at the same level as the less flexible ones. 
\\
Coal can only bid at one price level. By setting this price on the electricity bus, its bid could propagate to the heat bus through conversion technologies. This would allow gas CHP to set its bid at the same level as the coal bid. A similar effect can be obtained for storage technologies. Another example is the charging of BEVs. Their ability to shift load across hours throughout the year allows them to strategically place bids exactly at the market margin, while keeping volumes low enough to avoid triggering the next higher clearing price. To account for this variability in bids, we rank the technologies according to the variance of their bid/ask prices. The higher the variance, the more flexible the technology and the less likely it is to be a price setter. If there are multiple candidates, we select the one with the lowest variance. 

\subsection{Results}
\label{app:infos:results}

\subsubsection{Bid and ask behaviour of different technologies}
\label{app:infos:results:bid-ask}
\begin{figure}[!h]
    \centering
    \footnotesize
    (a) Ask prices of electricity generating technologies in 2025 and 2045 \\
    \includegraphics[width=0.8\textwidth]{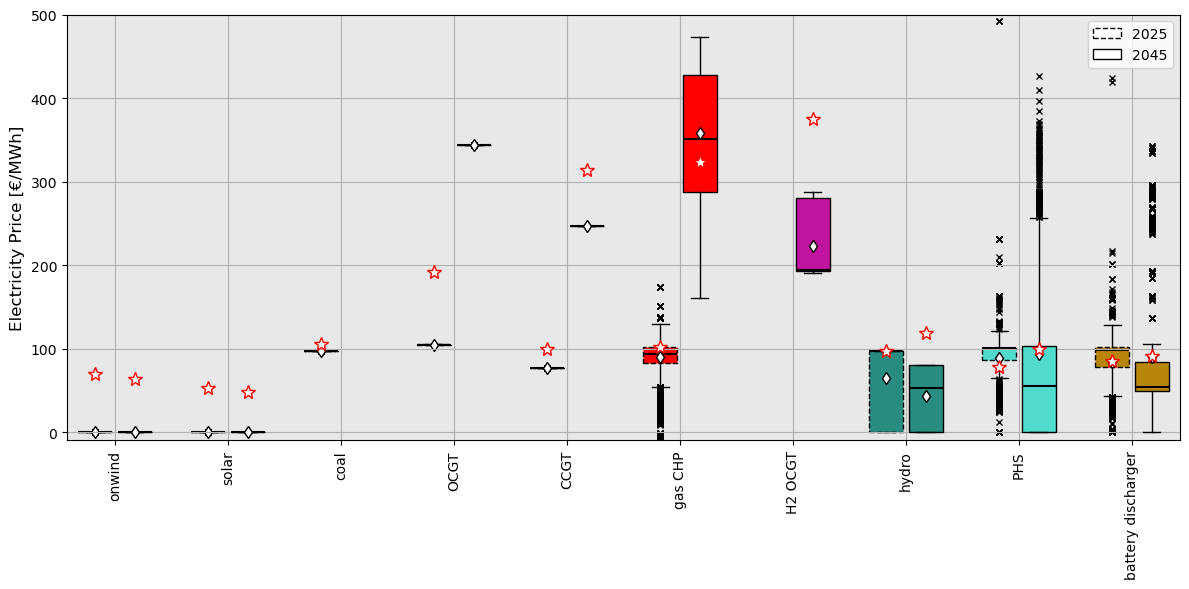} \\
    (b) Bid prices of electricity consuming technologies in 2025 and 2045 \\
    \includegraphics[width=0.8\textwidth]{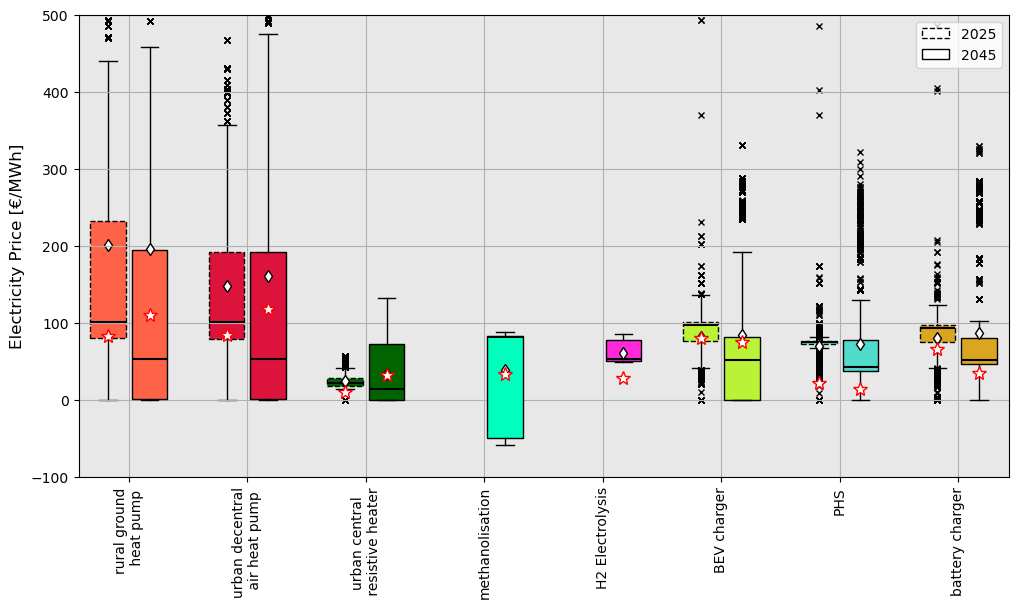} \\
    \caption{\textbf{Electricity bid and ask in 2025 and 2045.}
        The figures show the complete set of electricity ask prices (upper plot) and bid prices (lower plot) for the years 2025 and 2045. Red stars indicate the market values and purchasing prices (generation-weighted). Black lines indicate data median and black diamonds indicate data mean. For better visualisation the y-axis is limited.}
    \label{fig:bid-ask}
\end{figure}

In order to gain insights into the bid and ask behaviour of specific technologies and how they evolve towards a fully decarbonised system, we categorise the technologies and analyse their representative behaviour. Based on their ability to produce and consume electricity, and their flexibility in doing so, we obtain five distinctive categories. Fig.~\ref{fig:bid-ask} shows the bid and ask prices of the technologies in 2025 and 2045. The red stars indicate the market values and consumption-weighted purchasing prices.\\

Inflexible electricity producers are producers, whose bid volume is either highly dependent on weather conditions (VRES) or have high capital costs (coal, lignite, nuclear). They have limited ability to adjust production in response to market prices, so cannot exploit high prices. This results in a very small range of bid prices and rather low market values.\newline
Hydro, OCGT and CHP are more flexible either because they have lower capital costs (OCGT), can store energy (hydro) or can simultaneously produce heat (CHP). They have a wider range of bid prices and higher market values.\\
Heat pumps and resistive heaters are rather inflexible electricity consumers as they have fixed demand profiles and  limited storage capacity. However, their purchase prices are highly dependent on the heat price, which varies throughout the year. Bid prices for air heat pumps also depend on the ambient temperature, as this affects their efficiency. This results in a rather wide range of prices, especially at high levels. Compared to other electricity-consuming technologies, they consume at higher prices.\\
Electrolysis, methanolisation and BEV charger are flexible electricity consumers as they can either store their output ($H_2$) or do not have to follow a rigid demand profile. This results in relatively low electricity demand prices.\\ 
PHS and batteries are the most flexible technologies because they can dispatch, consume and store electricity. This capability is particularly important in 2045, when high VRES generation makes storage and flexibility indispensable (see Fig.~\ref{fig:bid-ask}). In 2025, the range of bid prices for PHS and battery dischargers is narrow, whereas in 2045, they cover almost all relevant price levels. Their market values are moderate and their purchase prices are relatively low. PHS has the lowest generation-weighted electricity purchasing price of 14~\euro/MWh, followed by electrolysis at 28~\euro/MWh.

\clearpage
\section{Tables and Figure}
\label{app:figures-tables}

\begin{table}[htbp]
\caption{Fuel price assumptions in \euro2020{}/MWh}
\begin{center}
\begin{threeparttable}
\begin{tabular}{lcccccc}
\toprule
\textbf{Fuel type} & \textbf{2020} & \textbf{2025} & \textbf{2030} & \textbf{2035} & \textbf{2040} & \textbf{2045} \\
\midrule
Oil                     & 52.91 & 52.91 & 52.91 & 52.91 & 52.91 & 52.91 \\
Hard coal               & 9.55 &  9.55 &  9.55 &  9.55 &  9.55 &  9.55 \\
Lignite                 & 3.30 &  3.30 &  3.30 &  3.30 &  3.30 &  3.30 \\
Gas                     & 24.57 & 24.57 & 24.57 & 24.57 & 24.57 & 24.57 \\
Solid biomass           & 13.65 & 13.65 & 13.65 & 13.65 & 13.65 & 13.65 \\
Solid biomass (import)  & 40.30 & 40.30 & 40.30 & 40.30 & 40.30 & 40.30 \\
\bottomrule
\end{tabular}
\begin{tablenotes}
\item[a] Assumptions based on data from the International and Danish Energy Agency \cite{schroder_current_2013} and \cite{pypsa_technology_data}.
\end{tablenotes}
\end{threeparttable}
\label{tab:fuel-costs}
\end{center}
\end{table}

\begin{figure}[htbp]
    \centering
    \includegraphics[width=0.5\linewidth]{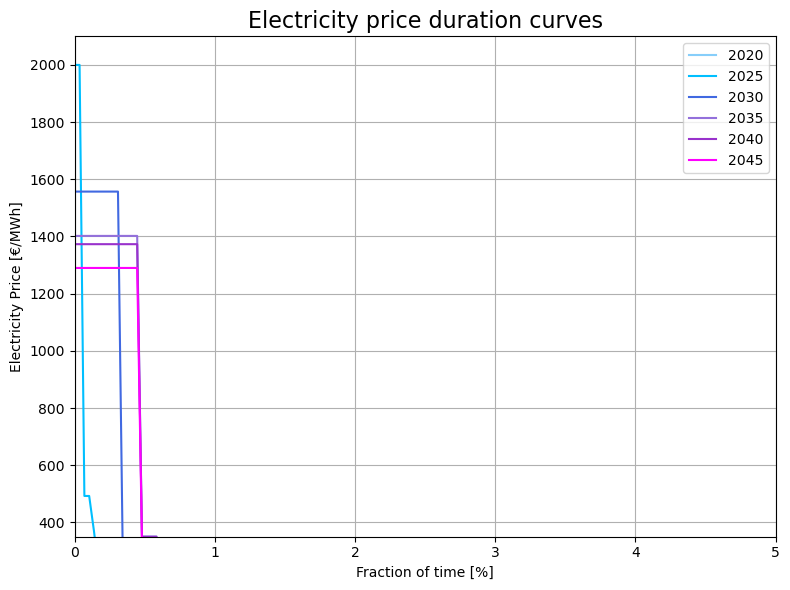}
    \caption{\textbf{Electricity price duration curves for different years representing different levels of CO$_2$ emissions.}
        The figure shows the development of the electricity price duration from the current system until a fully decarbonised system in 2045. Electricity prices above 350~\euro{}/MWh are shown.}
    \label{fig:elec-pdc-peak}
\end{figure}

\begin{figure}[h!]
    \centering
    \footnotesize

    \includegraphics[width=0.9\textwidth]{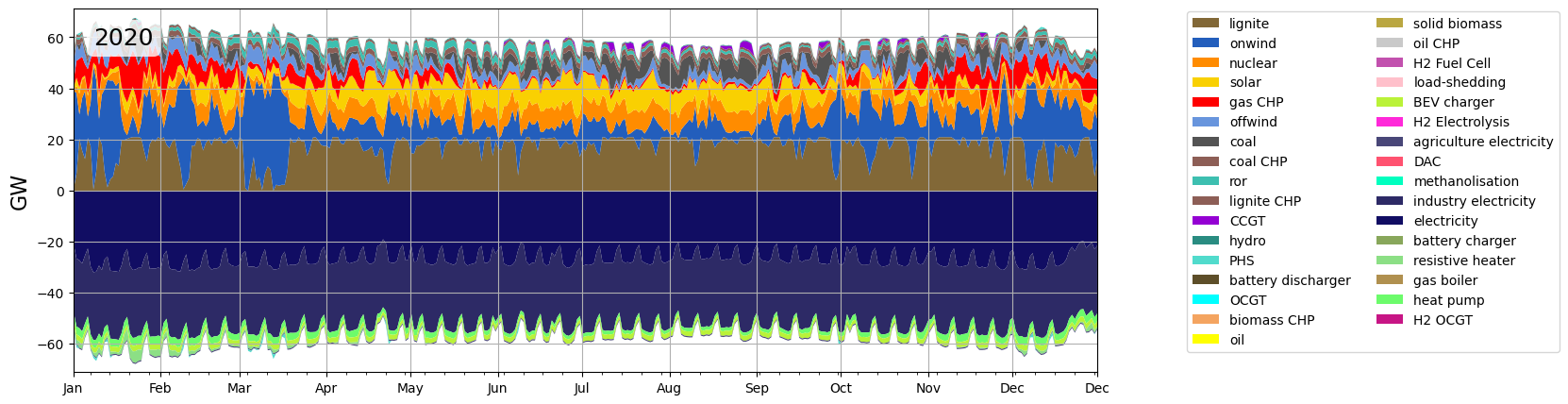} \\
    \includegraphics[width=0.9\textwidth]{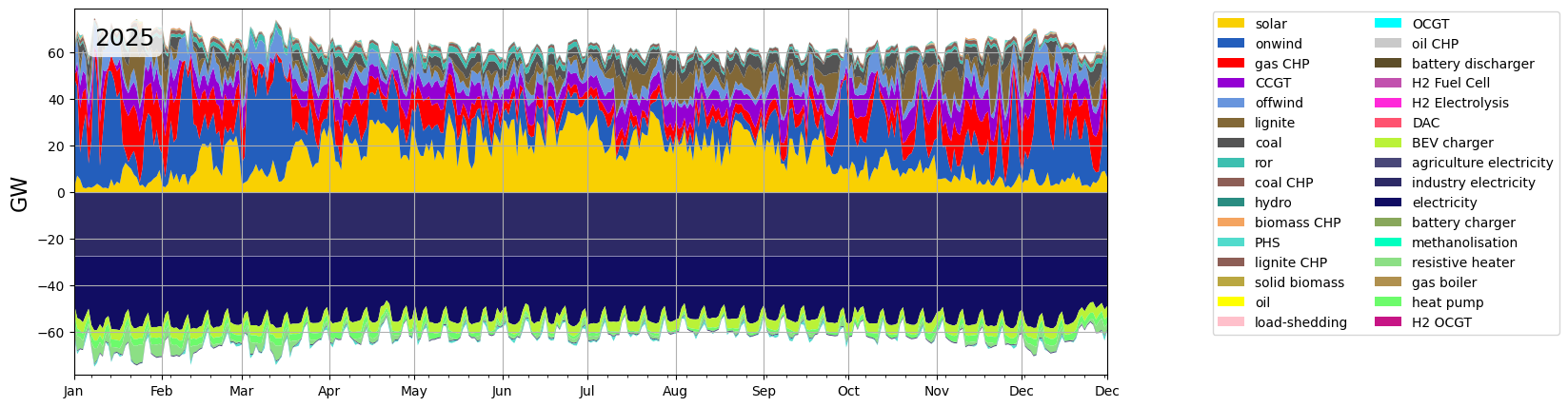} \\
    \includegraphics[width=0.9\textwidth]{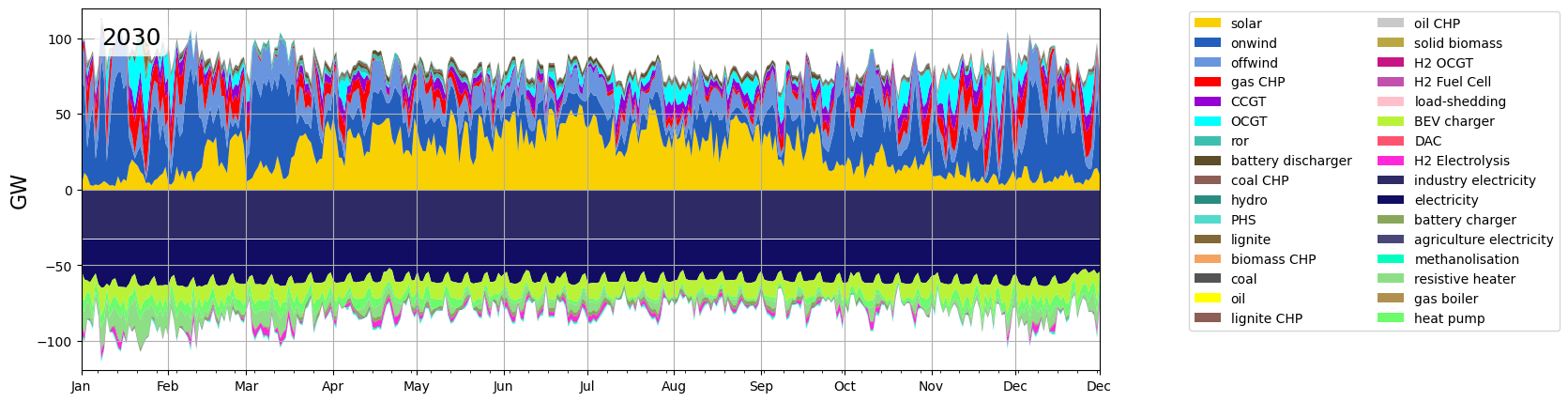} \\
    \includegraphics[width=0.9\textwidth]{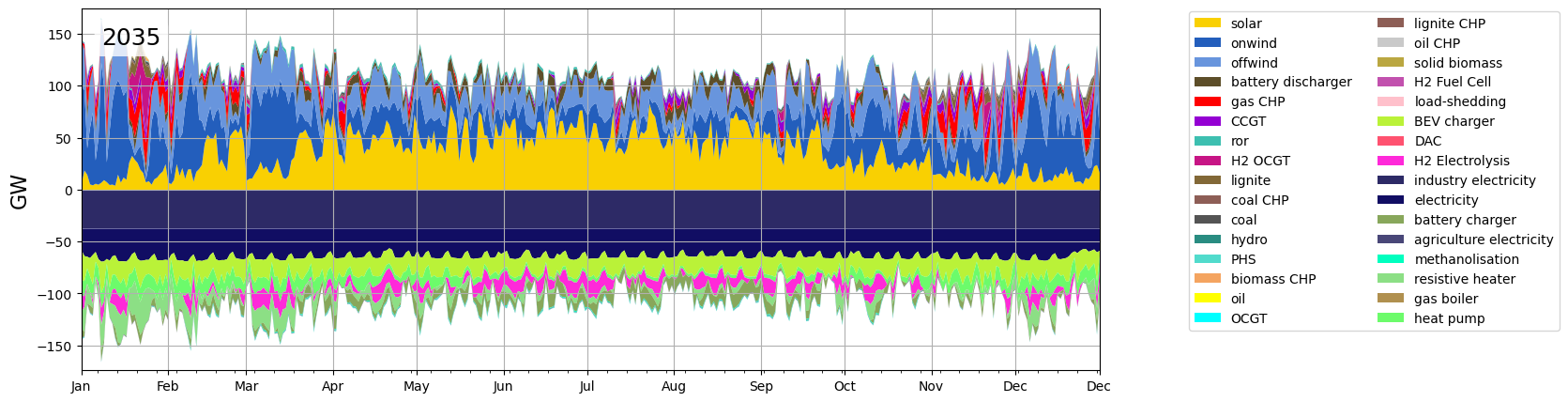} \\
    \includegraphics[width=0.9\textwidth]{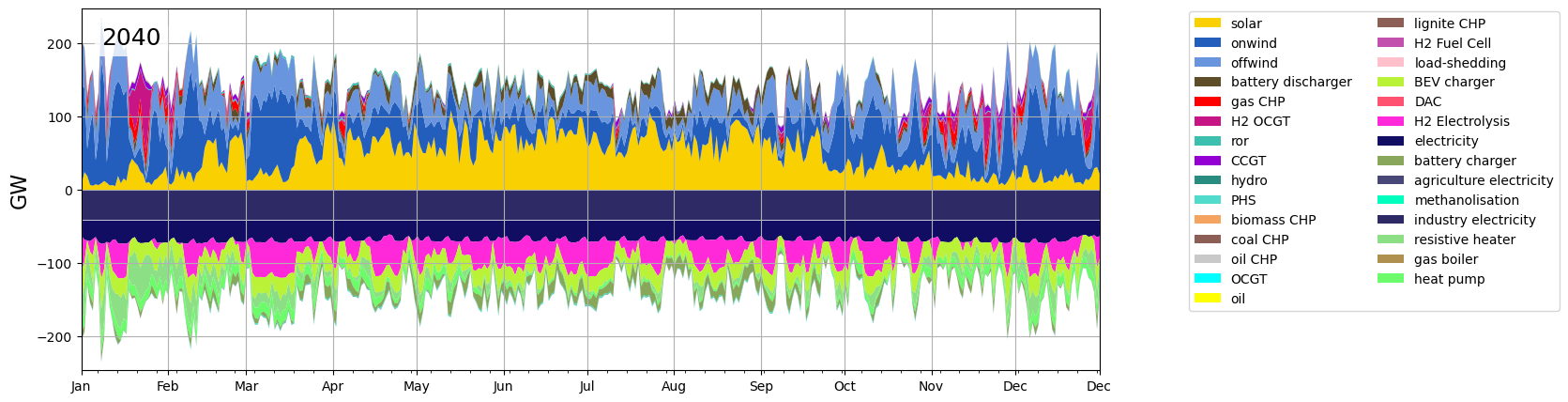} \\
    \includegraphics[width=0.9\textwidth]{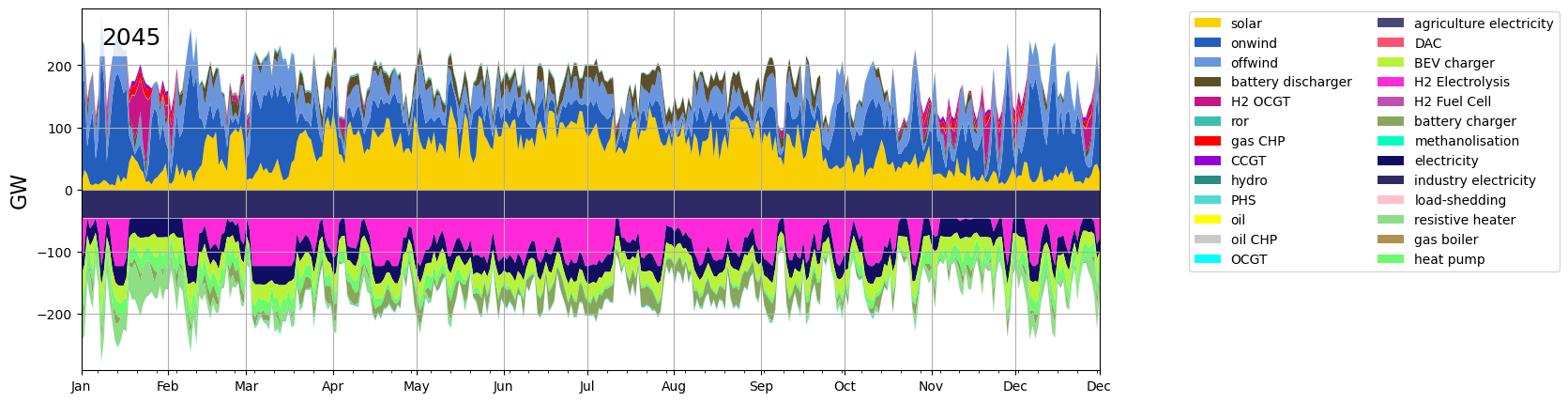} \\

    \caption{\textbf{Electricity Balances for 2020, 2025, 2030, 2035, 2040 and 2045.}
        }
    \label{fig:elec-balance-all}
\end{figure}

\begin{figure}[h!]
    \centering
    \footnotesize
    \includegraphics[width=1\textwidth]{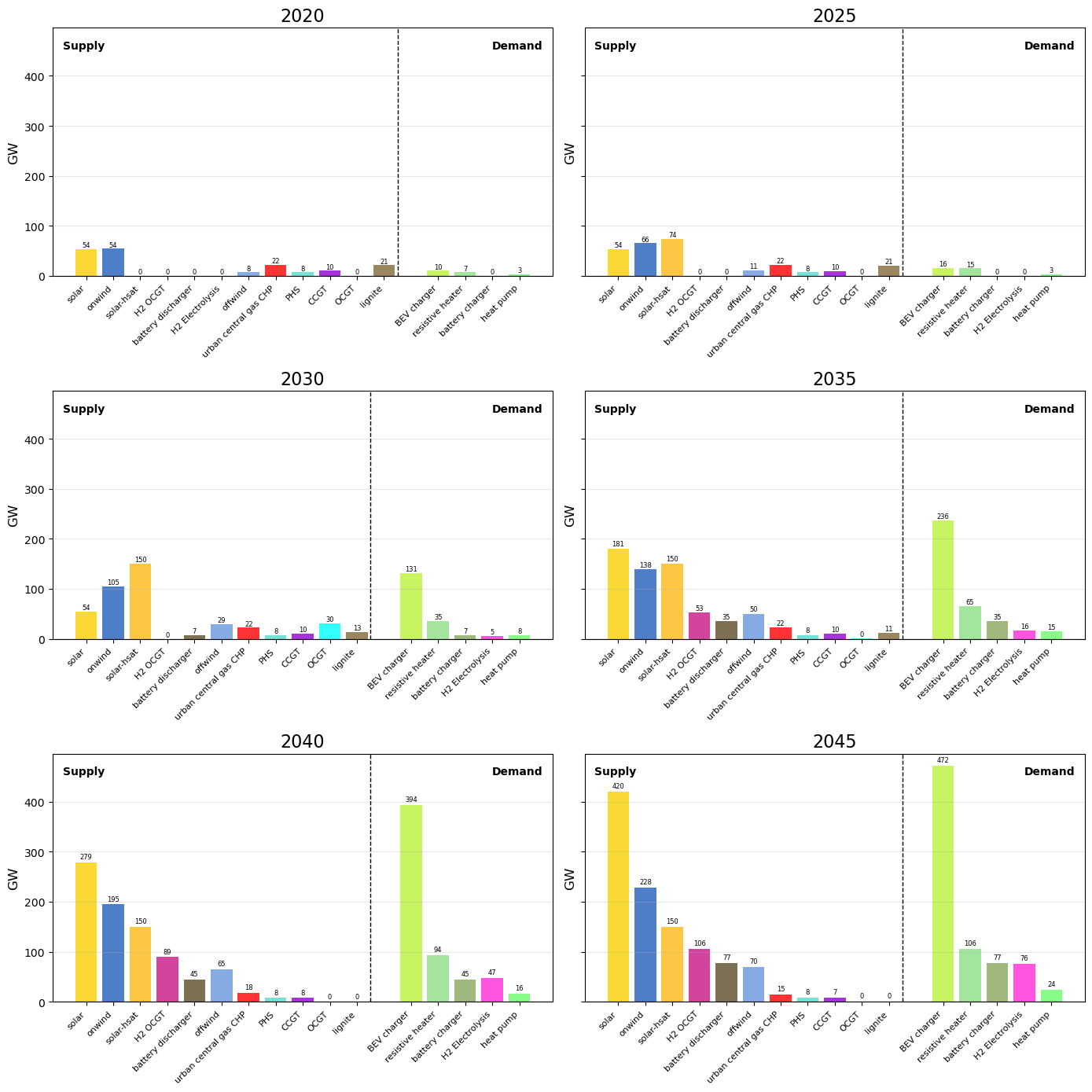} \\
    \caption{\textbf{Electricity generation and consumption capacity per technology from 2020-2045.}
        Installed electricity capacity per year (2020–2045) of the technologies with the highest overall capacity for all years.}
    \label{fig:elec-capa}
\end{figure}

\begin{figure}[h!]
    \centering
    \footnotesize
    \includegraphics[width=0.8\textwidth]{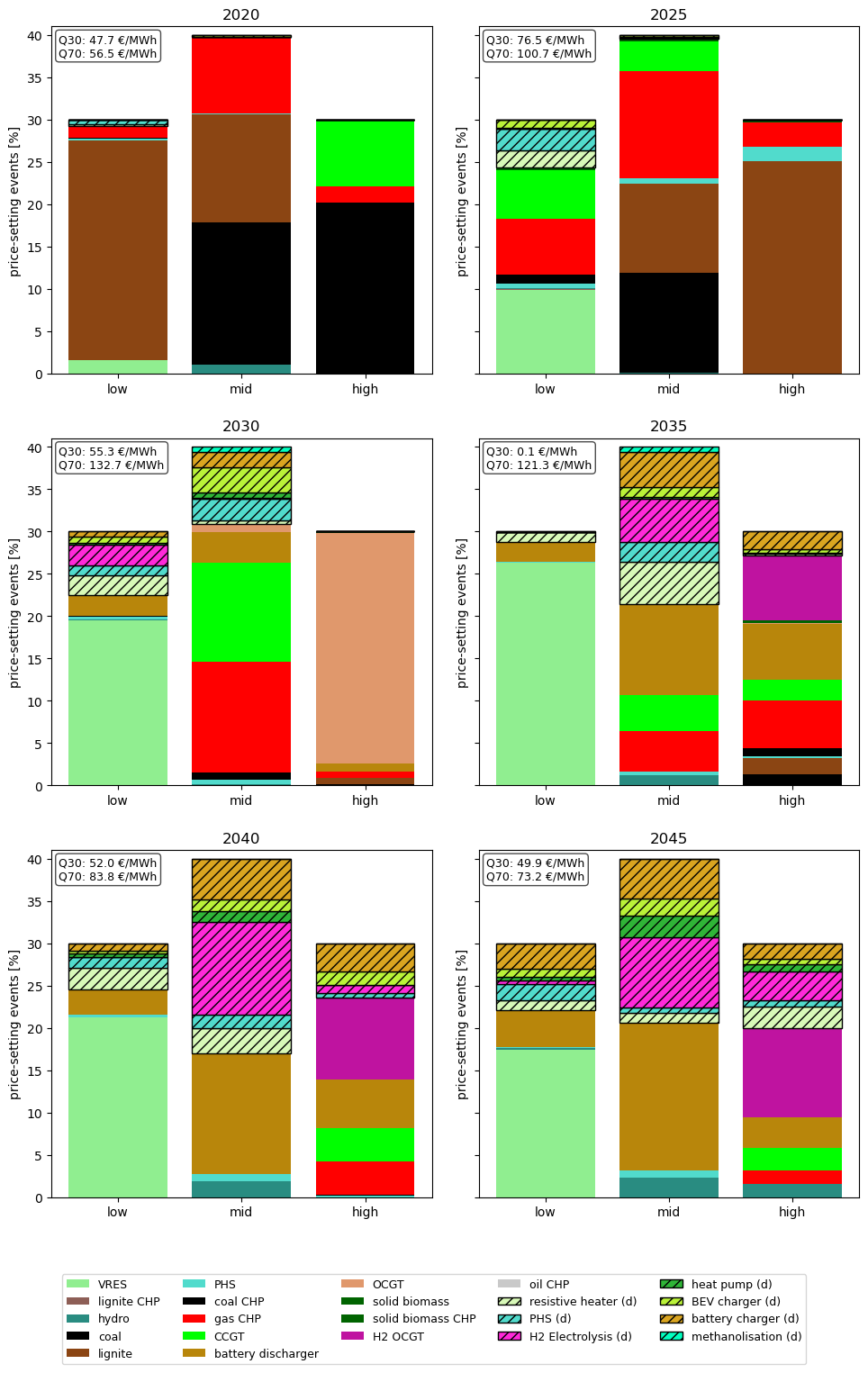} \\
    \caption{\textbf{Electricity price setter per technology and quantile from 2020-2045.}
        Price setting technologies in a high, average and low price situation. The high price is above the 0.7 quantile (Q70), the low price is below the 0.3 quantile (Q30), and the average price falls between these two thresholds.}
    \label{fig:price-setter-quantile}
\end{figure}

\begin{figure}[h!]
    \centering
    \footnotesize
    \includegraphics[width=0.9\textwidth]{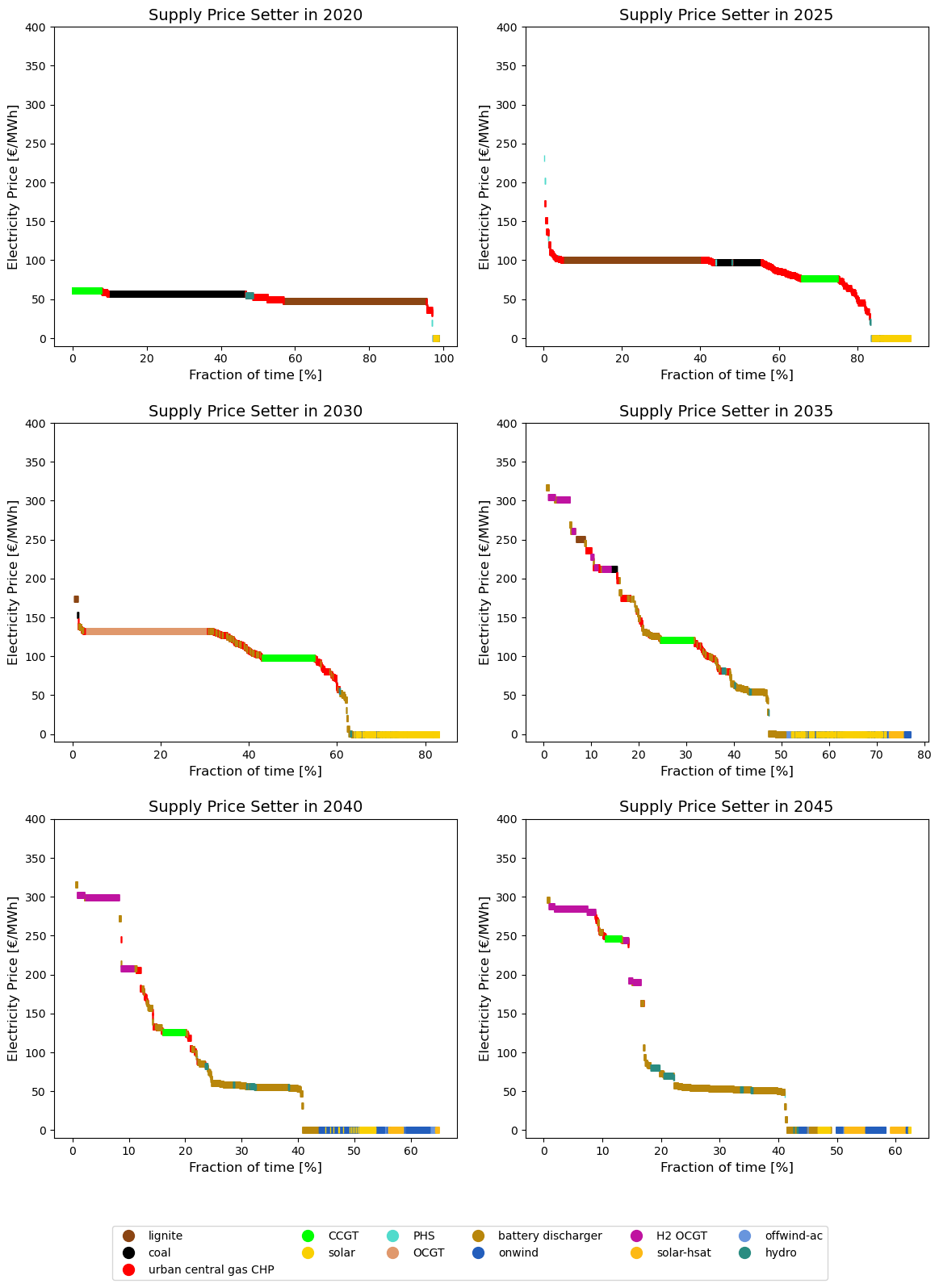} \\
    \caption{\textbf{Market clearing price duration curves for electricity (Supply).}
        The presented figures illustrate the price duration curves for electricity, with varying CO$_2$ constraints. The supply technology that sets the price is indicated.}
    \label{fig:market-clearing-pdc-supply}
\end{figure}

\begin{figure}[]
    \centering
    \footnotesize
    \includegraphics[width=0.9\textwidth]{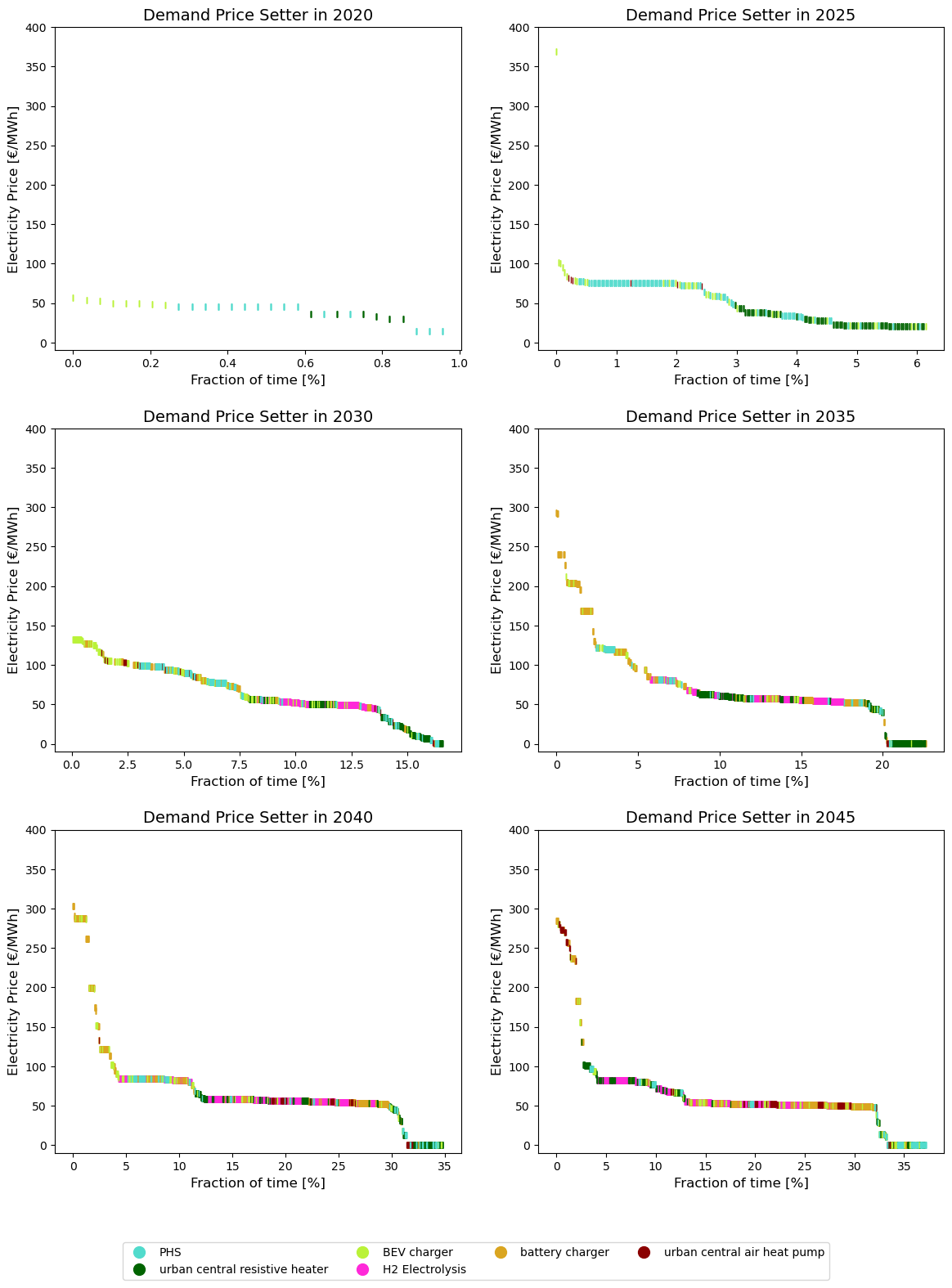} \\
    \caption{\textbf{Market clearing price duration curves for electricity (Demand).}
        The presented figures illustrate the price duration curves for electricity, with varying CO$_2$ constraints. The demand technology that sets the price is indicated.}
    \label{fig:market-clearing-pdc-demand}
\end{figure}

\begin{figure}[]
    \centering
    \footnotesize
    \includegraphics[width=0.9\textwidth]{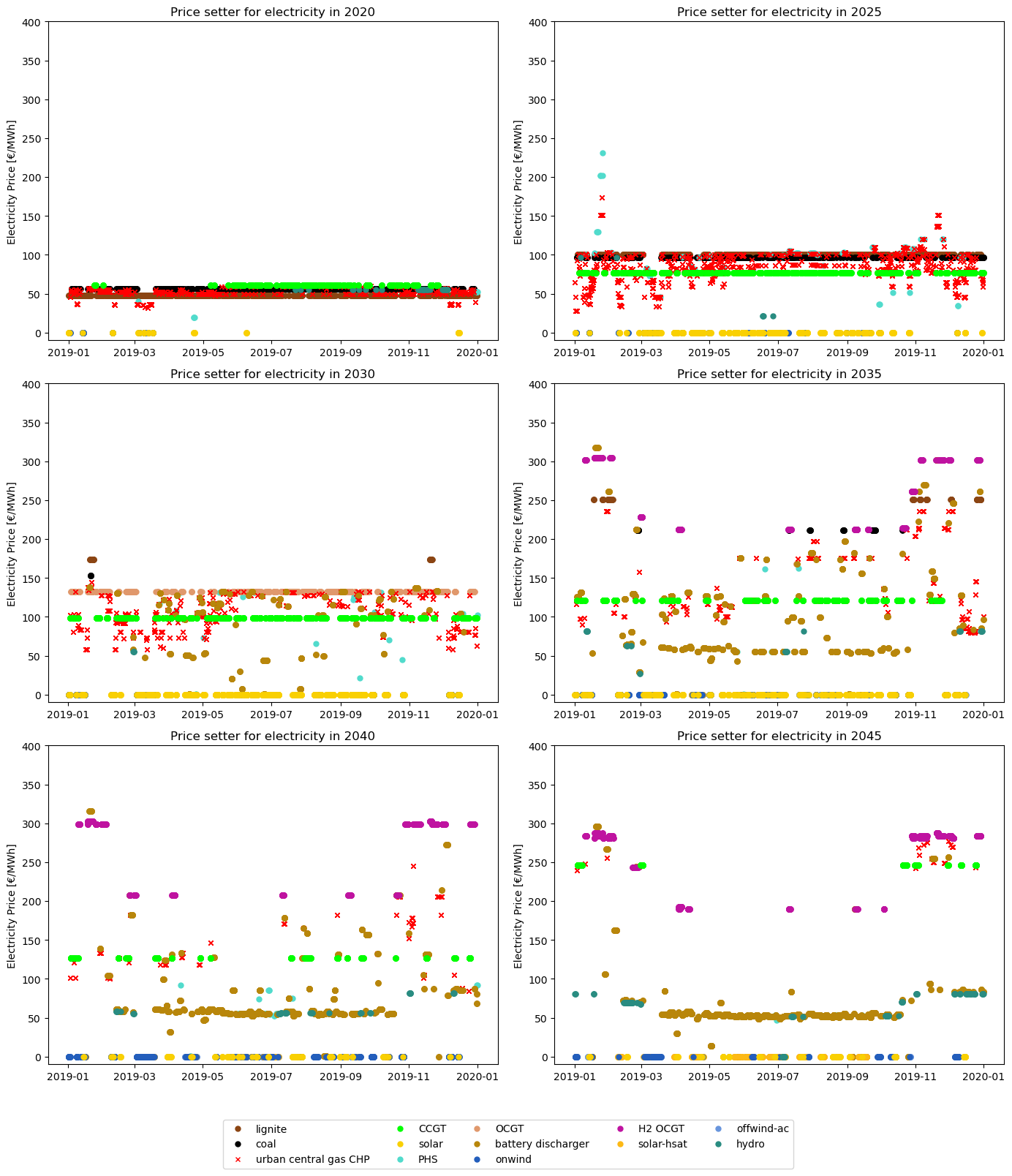} \\
    \caption{\textbf{Price setter temporal (Supply).}
        The presented figures illustrate the temporal price setting for electricity in different years representing different levels of CO$_2$ emissions. The supply technology that sets the price is indicated.}
    \label{fig:price-setting-temporal-supply}
\end{figure}

\begin{figure}[]
    \centering
    \footnotesize
    \includegraphics[width=0.9\textwidth]{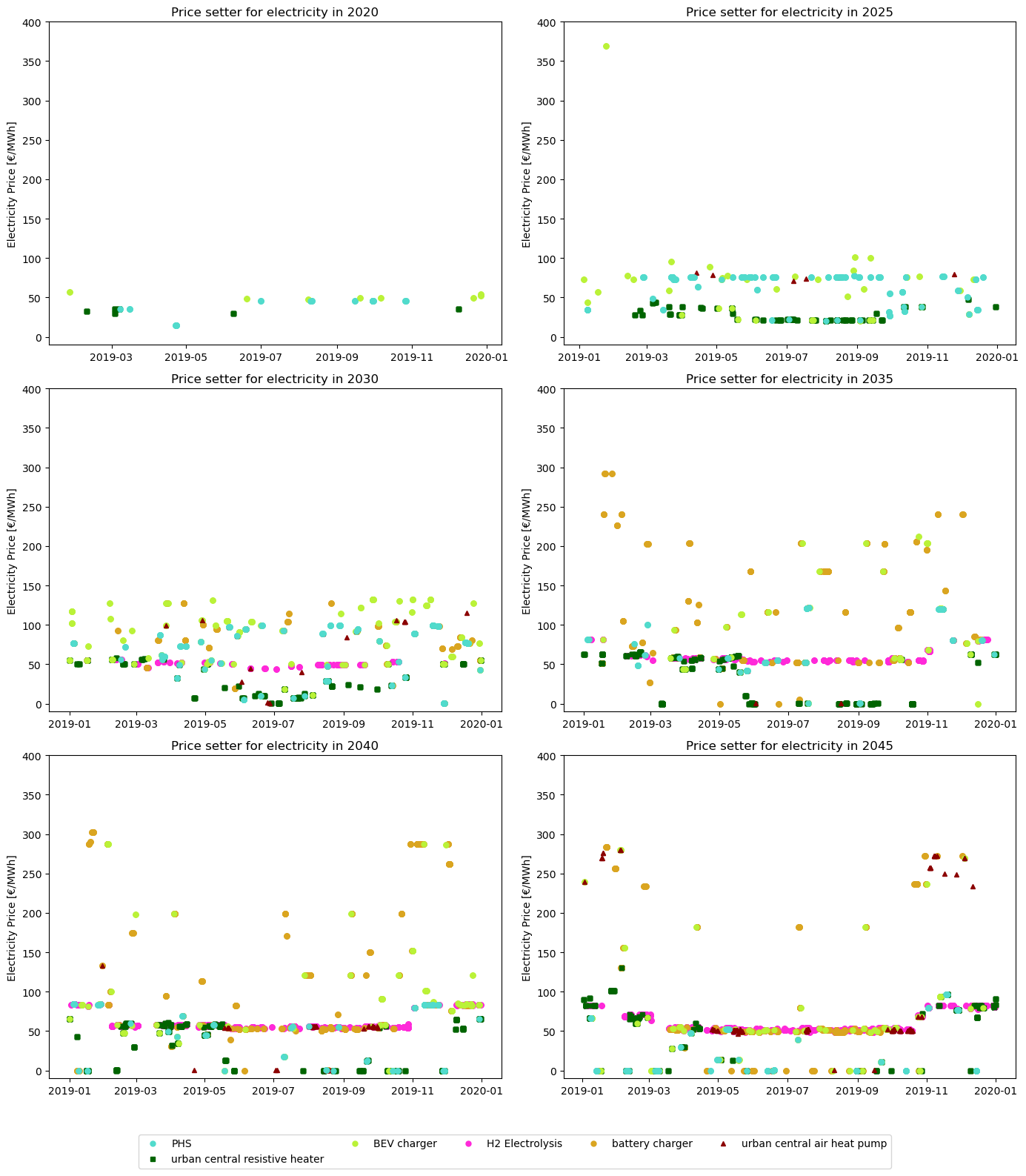} \\
    \caption{\textbf{Price setter temporal (Demand).}
        The presented figures illustrate the temporal price setting for electricity in different years representing different levels of CO$_2$ emissions. The demand technology that sets the price is indicated.}
    \label{fig:price-setting-temporal-demand}
\end{figure}

\begin{figure}[h!]
    \centering
    \footnotesize
    \includegraphics[width=1\textwidth]{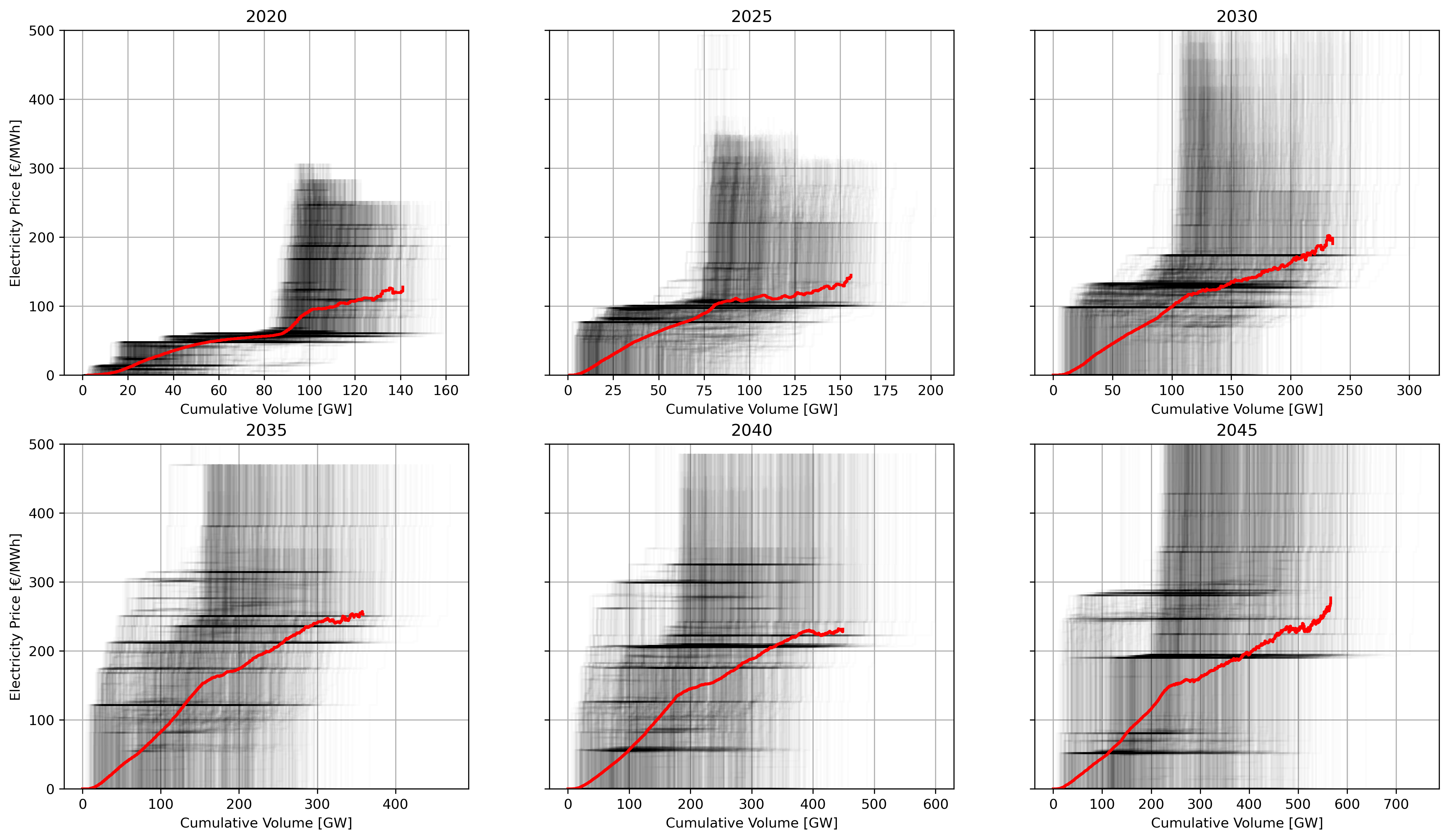} \\
    \caption{\textbf{Individual supply curves and aggregation.}
        Average electricity supply curves per year (2020–2045), with curves for all timesteps in black and the yearly average shown as a red line.}
    \label{fig:supply-curves}
\end{figure}

\begin{figure}[h!]
    \centering
    \footnotesize
    \includegraphics[width=1\textwidth]{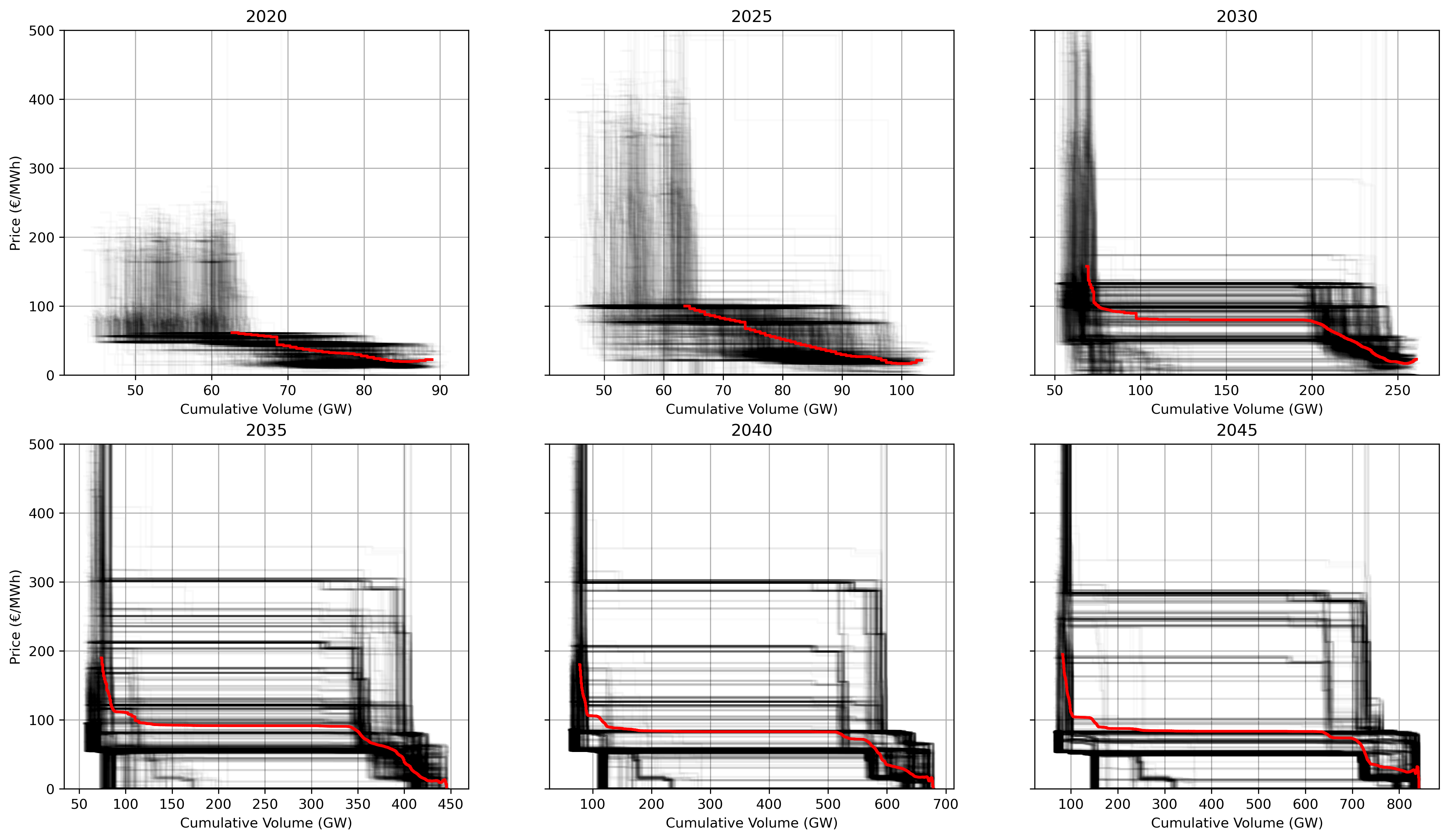} \\
    \caption{\textbf{Individual demand curves and aggregation.}
        Average electricity demand curves per year (2020–2045), with curves for all timesteps in black and the yearly average shown as a red line.}
    \label{fig:demand-curves}
\end{figure}

\begin{figure}[h!]
    \centering
    \footnotesize
    \includegraphics[width=1\textwidth]{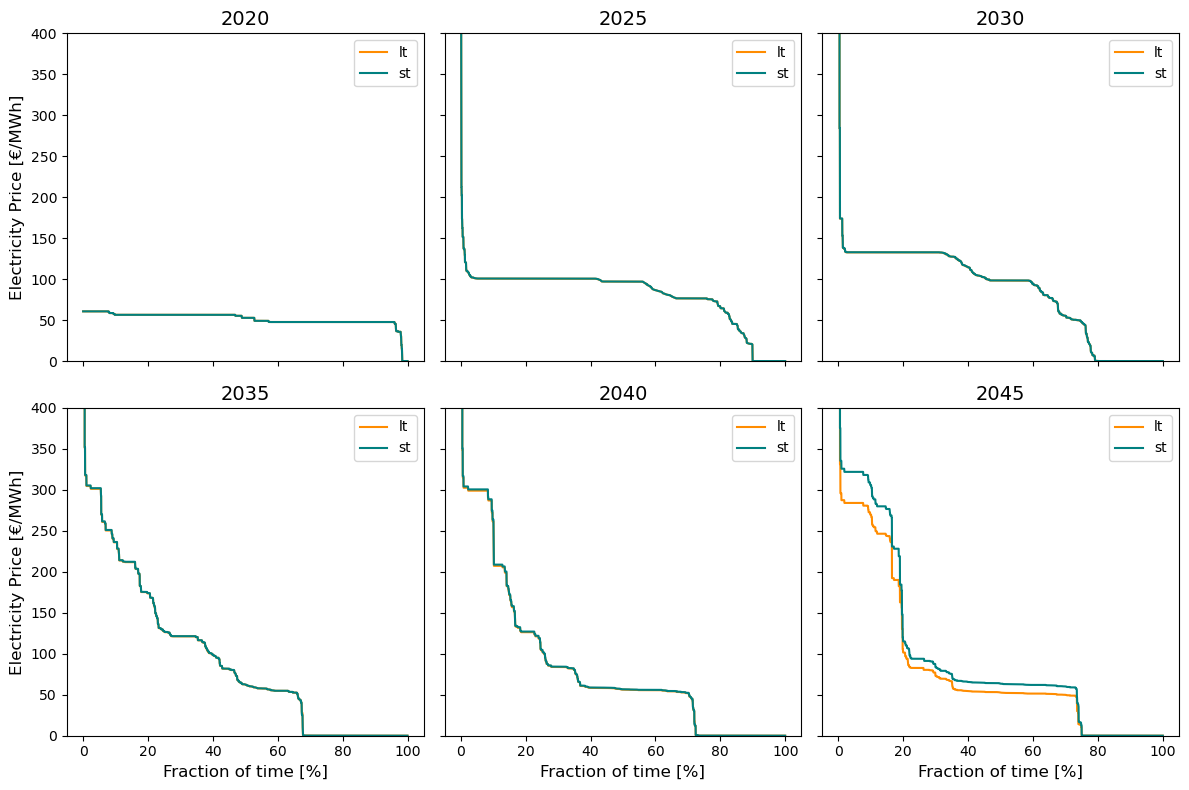} \\
    \caption{\textbf{Eletricity price duration curves for LT and ST model.}
        Comparison of the electricity price duration curves for the LT and ST model. The LT (long-term) model incorporates capacity and dispatch optimisation, while the ST (short-term) model only optimises dispatch.}
    \label{fig:pdc-st-lt}
\end{figure}

\begin{figure}[htb]
    \centering
    \footnotesize
    \begin{minipage}[t]{0.49\columnwidth}
        \centering
        (A) Development of price setting of electricity supply (ST model) \\
        \includegraphics[width=\linewidth]{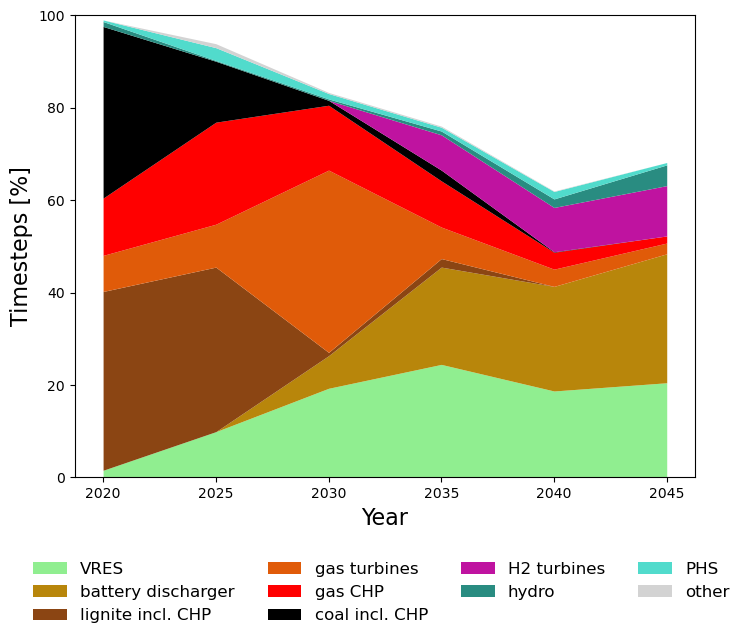}
    \end{minipage}
    \hfill
    \begin{minipage}[t]{0.49\columnwidth}
        \centering
        (B) Development of price setting of electricity demand (ST model) \\
        \includegraphics[width=\linewidth]{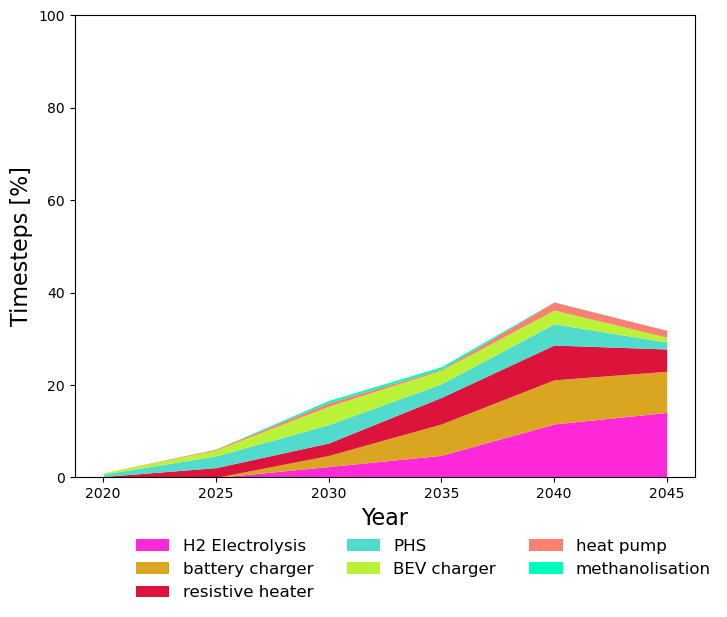}
    \end{minipage}
    \caption{\textbf{Evolution of price-setting technologies in the ST model.}
    The figure shows the number of timesteps at which an electricity supply or demand technology is price-setting in a specific year in the ST model. Technologies with fewer than 10 occurrences in a year are omitted.}
    \label{fig:price-setting-evo-st}
\end{figure}

\end{appendices} 

\end{document}